%% file: main.tex
\documentclass[journal]{IEEEtran}
\usepackage{amsmath}
\usepackage{amsfonts}
\usepackage{amssymb}
\usepackage{amsthm}
\usepackage{algorithmic}
\usepackage{algorithm}
\usepackage{array}
\usepackage{subcaption}
\usepackage{caption}
\usepackage[font=small,labelfont=bf]{caption}
\usepackage{textcomp}
\usepackage{stfloats}
\usepackage{xurl}
\usepackage{verbatim}
\usepackage{graphicx}
\usepackage{color}
\usepackage{cite}
\usepackage{multirow}
\usepackage{booktabs}
\usepackage{bm}
\usepackage{setspace}
\usepackage{arydshln}
\usepackage{flushend}
\usepackage{comment}
\usepackage{acronym}
\usepackage{threeparttable}
\usepackage{graphicx}
\usepackage{siunitx}
\usepackage{makecell}
\usepackage[colorlinks=true, linkcolor=blue, citecolor=blue, urlcolor=blue]{hyperref}

\input{abbr}
\begin{document}
\title{Robust Wildfire Forecasting under Partial Observability: From Reconstruction to Prediction}

\author{Chen Yang, Mehdi Zafari,~\IEEEmembership{Graduate Student Member,~IEEE,} Ziheng Duan,\\ and A.~Lee Swindlehurst,~\IEEEmembership{Life Fellow,~IEEE} %
\thanks{C. Yang, M. Zafari and A. Lee Swindlehurst are with the Department of Electrical Engineering and Computer Science, University of California, Irvine, CA 92697, USA (e-mail: \{cyang27, mzafarid, swindle\}@uci.edu).}%
\thanks{Z. Duan is with the Department of Computer Science, University of California, Irvine, CA 92697, USA (e-mail: zihend1@uci.edu).}%
}


\maketitle

\begin{abstract}
Satellite-derived fire observations are the primary input for learning-based wildfire spread prediction, yet they are inherently incomplete due to cloud cover, smoke obscuration, and sensor artifacts. This partial observability introduces a domain gap between the clean data used to train forecasting models and the degraded inputs encountered during deployment, often leading to unreliable predictions.
To address this challenge, we formulate wildfire forecasting under partial observability using a two-stage probabilistic framework that decouples observation recovery from spatiotemporal prediction. Stage-I reconstructs plausible fire maps from corrupted observations via conditional inpainting, while Stage-II models wildfire dynamics on the recovered sequences using a spatiotemporal forecasting network.
We consider four network architectures for the reconstruction module---a Residual U-Net (MaskUNet), a Conditional VAE (MaskCVAE), a cross-attention Vision Transformer (MaskViT), and a discrete diffusion model (MaskD3PM)---spanning CNN-based, latent-variable, attention-based, and diffusion-based approaches. We evaluate the performance of the two-stage approach on the \acf{wsts} dataset under various settings including pixel-wise and block-wise masking, eight corruption levels (10\%--80\%), four fire scenarios, and leave-one-year-out cross-validation. Results show that all learning-based recovery models substantially outperform non-learning baselines, with MaskCVAE and MaskUNet achieving the strongest overall performance. Importantly, inserting the reconstruction stage before forecasting significantly mitigates the domain gap, restoring next-day prediction accuracy to near-clean-input levels even under severe information loss.
\end{abstract}


\section{Introduction}
\label{sec:intro}

\subsection{Background}
\IEEEPARstart{I}{n} recent years, the escalating frequency and severity of natural disasters have imposed staggering costs on economies, ecosystems, and human well-being worldwide. Among these hazards, wildfires stand out as one of the most destructive, inflicting damage that spans multiple dimensions of society. Economically, wildfires have caused tens of billions of dollars in annual economic damages globally in recent years~\cite{sullivan2022spreading}, with wildfire smoke-attributable mortality imposing an additional cumulative cost of 160~billion~USD over the period of 2006--2020~\cite{law2025anthropogenic}. From an ecological standpoint, the 2019--2020 Australian ``Black Summer'' burned over 30 million hectares of vegetation and killed or displaced an estimated 3 billion terrestrial vertebrates~\cite{dickman2021ecological}, devastating critical habitats including more than 80\% of the World Heritage Blue Mountains National Park~\cite{reiner2024_black_summer_tourism}. Wildfires also exact a severe toll on human life and public health. The January 2025 Los Angeles wildfire complex destroyed over 16{,}000 structures, claimed at least 31 lives directly, and contributed to an estimated 440 excess deaths from smoke exposure and healthcare disruption~\cite{paglino2025excess}.

Wildfire detection and forecasting fundamentally rely on observational infrastructure. Ground-based sensor networks (e.g., lookout towers, weather stations, and in situ fuel-moisture probes) provide high temporal resolution, physically grounded measurements of key variables such as temperature, humidity, wind speed, and fuel moisture content~\cite{chan2024survey}. However, their spatial coverage is confined to instrumented sites, limiting their ability to monitor large-scale fires across heterogeneous terrain. To overcome these spatial limitations, airborne and spaceborne remote sensing platforms have become the backbone of modern wildfire monitoring. Aircraft-mounted thermal sensors and, more prominently, satellite systems such as \ac{modis}~\cite{justice2002modis} and \ac{viirs}~\cite{Schroeder2014VIIRS} provide global, multi-spectral observations at regular intervals, enabling the detection of active fire hotspots, estimation of fire radiative power, and mapping of burned areas at continental to global scales. The synoptic view and temporal consistency of satellite observations have made them indispensable for operational fire management agencies worldwide. More recently, \acp{uav} and ground-based robotic platforms have emerged as complementary tools for fine-grained, localized fire observation~\cite{bouguettaya2022review}. Equipped with thermal and multispectral cameras, these systems can provide centimeter-level spatial resolution and flexible deployment in specific areas of interest. However, their limited endurance, small coverage footprint, and dependence on line-of-sight communication constrain their utility primarily to tactical, short-duration monitoring of individual fire events rather than systematic, long-term surveillance.

In the data-driven era of wildfire research, satellite remote sensing has become the dominant observational resource for fire monitoring, owing to several practical and physical advantages. First, polar-orbiting and geostationary platforms offer near-global coverage with regular revisit schedules, sub-daily for geostationary sensors and daily to multi-day for polar orbiters, supporting systematic, temporally consistent monitoring over extended periods~\cite{savtchenko2004terra,schmit2017closer}. Second, multi-spectral observations provide complementary fire-relevant signals: mid-/thermal-infrared (MIR/TIR) measurements enable thermal anomaly and active-fire detection, visible/near-infrared (VIS/NIR; VNIR) reflectance characterizes vegetation condition, shortwave infrared (SWIR) bands are sensitive to vegetation and fuel water content through liquid-water absorption features, and microwave (MW) observations provide additional information on moisture conditions, together capturing key controls on fire activity and spread~\cite{chuvieco2020satellite}. Third, mature downstream products, including \ac{viirs} active-fire detections~\cite{Schroeder2014VIIRS}, \ac{modis} burned-area maps~\cite{justice2002modis}, and gridded meteorological reanalysis fields~\cite{Abatzoglou2013GriddedMet}, provide readily usable, spatiotemporally harmonized inputs that integrate naturally into modern machine learning approaches.

\subsection{Prior Art}
Several satellite-derived datasets have been developed to support wildfire analysis. Huot et al.~\cite{huot2022next} introduced the \ac{ndws} dataset, which combines nearly a decade of \ac{viirs} active fire detections across the contiguous United States with eleven inter-related environmental variables, including topography, vegetation index, weather, and drought indicators, at daily temporal resolution, targeting single-day fire spread prediction. This was extended by Gerard et al. in~\cite{Gerard2023WildfireSpreadTS} with \ac{wsts}, a multi-modal time-dataset that preserves temporal ordering across consecutive days, enabling the study of sequential fire dynamics over multiple days. More recently, Zhao et al.~\cite{zhao2025ts} proposed TS-SatFire, a multi-task dataset built from \ac{viirs} image time-series that unifies active fire detection, burned area mapping, and fire progression prediction within a single benchmark. Beyond \ac{viirs}-centric resources, Sykas et al.~\cite{sykas2023eo4wildfires} released EO4WildFires, which integrates Sentinel-2 multispectral imagery, Sentinel-1 \ac{sar} data, and meteorological variables across 45 countries for wildfire severity estimation, while Kondylatos et al.~\cite{kondylatos2023mesogeos} oversaw collection of the Mesogeos dataset covering the Mediterranean basin using \ac{modis} products and ERA5 reanalysis for wildfire prediction. Collectively, these efforts highlight the central role of satellite observations in enabling data-driven wildfire modeling.

Enabled by the growing availability of these multi-modal satellite datasets, learning-based methods for wildfire spread prediction have attracted considerable research attention~\cite{ghali2023deep,andrianarivony2024machine}. Early efforts trained \acp{cnn} on remote-sensing inputs to learn the complex nonlinear relationships among topography, vegetation, meteorology, and fire behavior, enabling next-day forecasts of active fire extent~\cite{radke2019firecast,huot2022next}. Subsequent work incorporated temporal modeling: spatiotemporal architectures such as \ac{convlstm}~\cite{shi2015convlstm} and \ac{utae}~\cite{garnot2021panoptic}, when trained on multi-day fire sequences, were shown to outperform single-snapshot baselines by capturing sequential spread dynamics~\cite{Gerard2023WildfireSpreadTS}. More recently, Transformer-based models~\cite{li2024wildfirevit} and U-Net variants~\cite{khennou2023improving} have been applied to satellite-derived fire sequences, achieving state-of-the-art performance in burned-area mapping and fire progression forecasting~\cite{ghali2023deep,zhao2025ts}. When provided with sufficiently rich and accurate training data, such models have demonstrated promising results, surpassing the predictive accuracy of conventional physics-based fire simulators~\cite{andrianarivony2024machine}.

\subsection{Contributions}

Despite these advances, a critical yet under-explored challenge persists: satellite-based fire observations are inherently only partially observed~\cite{schroeder2008quantifying, giglio2016collection}. Thermal anomaly detections can be missed or corrupted by cloud cover, dense smoke plumes generated by the fire itself, unfavorable satellite viewing angles, and sensor artifacts, leading to an observed fire state that is rarely a faithful representation of the true underlying fire extent. However, the datasets surveyed above are predominantly derived from selectively filtered, high-quality acquisitions, and the models trained on them implicitly assume that complete and reliable fire maps are available at inference time~\cite{radke2019firecast, huot2022next, Gerard2023WildfireSpreadTS}. This assumption introduces a fundamental gap: when deployed in operational settings where observations are inevitably degraded, the distributional shift between clean training data and corrupted test inputs can cause severe performance degradation, producing unreliable predictions when accurate forecasts are most needed. Therefore, addressing this partial observability represents a key open problem for robust, real-world wildfire forecasting.

To address this challenge, we propose a two-stage framework that explicitly decouples the problems of \emph{reconstruction} and \emph{forecasting}. Rather than expecting a single model to simultaneously compensate for missing data and predict future fire evolution, we first recover a plausible complete observation sequence from the corrupted inputs (Stage-I: Morphological Reconstruction), and then apply a forecasting model to the recovered sequence (Stage-II: Spatiotemporal Prediction). This decomposition is grounded in a principled probabilistic factorization of the predictive distribution (detailed in Section~\ref{subsec:problem_formulation}), and isolates the partial observability challenge into a dedicated generative module while allowing the forecaster to operate within its valid training domain.

The main contributions of this work are as follows:
\begin{itemize}
    \item Unlike prior studies that rely on the unrealistic assumption of fully observable training data~\cite{huot2022next, Gerard2023WildfireSpreadTS}, we explicitly confront the inherent partial observability in real-world wildfire records. To address this critical gap, we formulate wildfire forecasting as a two-stage problem, deriving a rigorous probabilistic justification for decoupling historical reconstruction from future prediction.
    \item We pioneer a comprehensive methodology for fire map reconstruction under partial observability. Rather than relying on a single architecture, we formulate four complementary paradigms for Stage-I: a Residual U-Net (MaskUNet)~\cite{Ronneberger2015UNet, he2016resnet}, a Conditional Variational Autoencoder (MaskCVAE)~\cite{kingma2013vae, sohn2015cvae}, a cross-attention Transformer (MaskViT)~\cite{vaswani2017attention, dosovitskiy2020vit}, and a discrete diffusion model (MaskD3PM)~\cite{ho2020denoising, austin2021structured}. This investigation uncovers the core architectural trade-offs among CNN-based, latent-variable, attention-based, and diffusion-based strategies in modeling sparse binary dynamics.
    \item We design a multi-dimensional benchmarking protocol on the \ac{wsts} dataset~\cite{Gerard2023WildfireSpreadTS}, built upon a dynamic spatial-focusing strategy. By dynamically cropping around active fire regions, we mitigate the severe class imbalance inherent in wildfire data (where fire pixels typically constitute less than 5\% of each frame) and force the models to learn fine-grained propagation behaviors within actual burning zones. Under this data setup, our protocol evaluates the paradigms across 2 corruption mechanisms, pixel-wise and block-wise masking, 8 severity levels, 4 fire scenarios, and a leave-one-year-out cross-validation scheme spanning 4 years of data.
    \item We show that our Stage-I recovery effectively neutralizes the catastrophic domain gap induced by missing data, restoring downstream forecasting capabilities to near-oracle parity, exhibiting resilience even under an extreme 80\% information deficit.
\end{itemize}

\subsection{Organization and Notation}
The remainder of the paper is organized as follows. Section~\ref{sec:method} details our proposed two-stage forecasting framework for handling partial observability, including the problem formulation, Stage-I morphological reconstruction module, and Stage-II spatiotemporal prediction framework. Section~\ref{sec:experiments} presents the experiments, encompassing multi-modal dataset preprocessing, benchmarking protocol, and a comprehensive evaluation of the recovery and prediction performance of different models under various missing data scenarios. Section~\ref{sec:conclusion} concludes the paper and discusses future research directions.

Regarding the mathematical notation used in this paper, we employ uppercase bold letters (e.g., $\mathbf{X}_t, \mathbf{E}_t, \mathbf{F}_t$) to represent high-dimensional tensors or matrices such as system states, environmental contexts, and fire characteristics. Calligraphic letters (e.g., $\mathcal{H}_t$) are specifically used to denote the set of historical sequences over multiple time steps. To distinguish the observational status of the data, variables with a tilde (e.g., $\tilde{\mathbf{X}}_\tau^{(\eta)}$) denote corrupted or partially observed data due to sensor limitations and other factors, while variables with a hat (e.g., $\hat{\mathbf{F}}_t$) represent the predicted results generated or reconstructed by the models. Additionally, the symbol $\odot$ is used to denote the element-wise multiplication (Hadamard product) of matrices, and lowercase Greek letters (e.g., $\tau, \eta$) are used to represent scalar parameters such as time steps and missing ratios. 
\begin{figure*}[t!]
    \centering
\includegraphics[width=\linewidth]{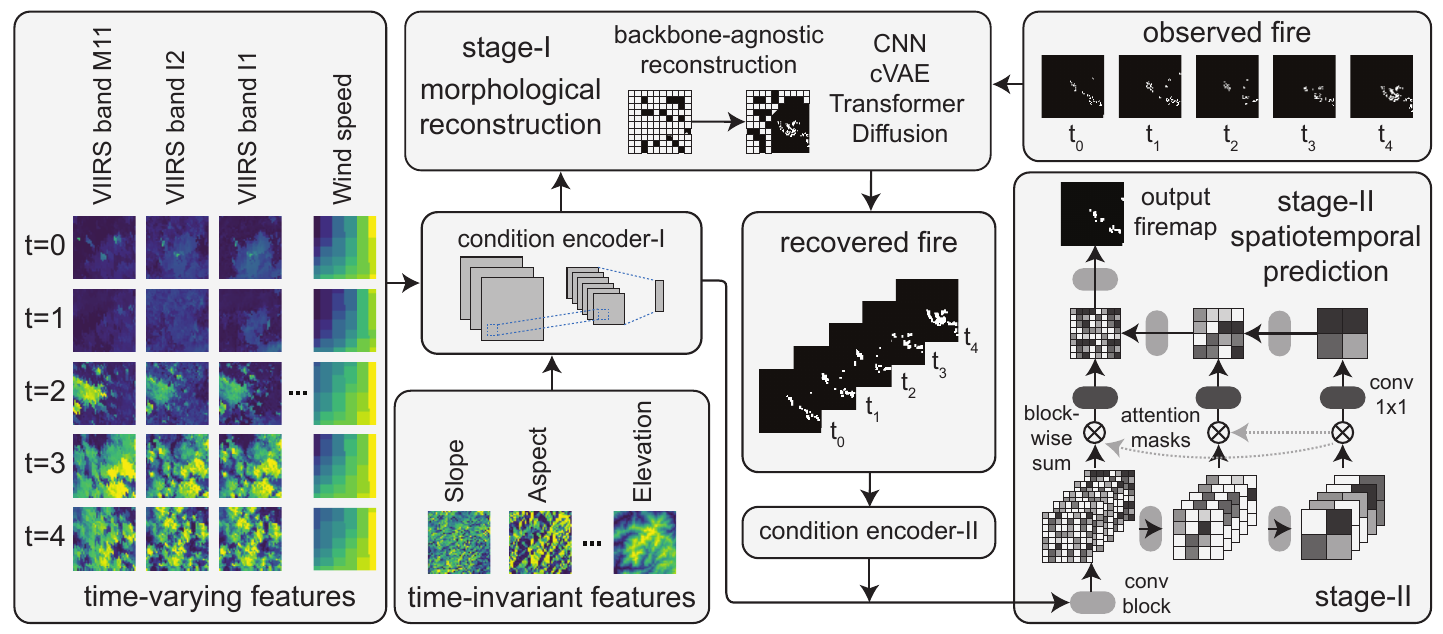}
    \caption{
    Overview of the proposed two-stage framework for wildfire forecasting under partial observability. Given multi-modal environmental observations together with partially observed fire maps, Stage-I performs \textit{morphological reconstruction} to recover a plausible complete fire history from corrupted inputs, explicitly addressing the domain gap introduced by missing satellite observations. The reconstruction module can be implemented with various different approaches, including CNN-based, latent generative, transformer, or discrete diffusion networks. The recovered sequence is then fed into Stage-II, a spatiotemporal prediction model that learns wildfire dynamics and produces the future fire map $\hat{\mathbf{F}}_t$. By decoupling observation recovery from temporal forecasting, the framework enables robust prediction under severe observation degradation.
    }
    \label{fig:framework}
\end{figure*}

\section{Methodology}
\label{sec:method}

We propose a two-stage framework for wildfire forecasting under partial observability, as illustrated in {Fig.~\ref{fig:framework}}. The framework consists of (i) a \textit{Morphological Reconstruction} stage that recovers corrupted fire observations, and (ii) a \textit{Spatiotemporal Prediction} stage that models future fire dynamics based on the reconstructed history. Detailed architectural settings for the models presented in this section are provided in Appendix~\ref{app:implementation}. An open-source Python implementation of the entire framework, including all configurations, seed numbers, and datasets are available at~\cite{githubrepo}.

\subsection{Problem Formulation}
\label{subsec:problem_formulation}
We formulate wildfire forecasting under partial observability as a spatiotemporal sequence prediction problem. The system state at day $t$ is defined by $\mathbf{X}_t = [\mathbf{E}_t ; \mathbf{F}_t]$, where $\mathbf{E}_t \in \mathbb{R}^{C_E \times H \times W}$ represents the multi-modal environmental context (e.g., meteorology, topography, vegetation) and $\mathbf{F}_t \in \{0,1\}^{H \times W}$ represents the binary active fire map ($1=$ fire present, $0=$ fire absent). The dimension $C_E$ represents the number of different channels of environmental features, while $H$ and $W$ denote the number of cells in the ground-level two-dimensional observation grid, respectively. Assuming an observation window over $T$ days, the true state sequence is denoted by $\mathcal{H}_{t-1} = \{\mathbf{X}_{t-T}, \dots, \mathbf{X}_{t-1}\}$. 

In a real-world deployment, the fire channel is only partially observed due to various factors such as smoke/cloud occlusions or sensor limitations. For each timestep $\tau \in \{t-T,\dots,t-1\}$, we define a binary corruption mask $\mathbf{M}_\tau^{(\eta)} \in \{0,1\}^{H \times W}$, where 0 indicates a cell with missing fire data. The variable $\eta$ indicates the percentage of observation cells that are missing, which is assumed to be constant over the observation window. The partially observed fire map is then constructed as $\tilde{\mathbf{F}}_\tau^{(\eta)} = \mathbf{F}_\tau \odot \mathbf{M}_\tau^{(\eta)}$, where $\odot$ denotes the element-wise (Hadamard) product. The environmental state $\mathbf{E}_t$ is assumed to be known without error. Consequently, the corrupted system state is represented as $\tilde{\mathbf{X}}_\tau^{(\eta)} = [\mathbf{E}_\tau ; \tilde{\mathbf{F}}_\tau^{(\eta)}] \in \mathbb{R}^{(C_E+1) \times H \times W}$, and the state sequence available for analysis is defined as $\tilde{\mathcal{H}}_{t-1} = \{\tilde{\mathbf{X}}_{t-T}^{(\eta)}, \dots, \tilde{\mathbf{X}}_{t-1}^{(\eta)}\}$.
The asymmetry in knowledge about $\mathbf{E}_\tau$ and $\tilde{\mathbf{F}}_\tau^{(\eta)}$ forces the learned model to perform cross-modal reasoning and synthesize sparse fire evidence with complete environmental priors, thereby validating its ability to infer missing dynamics based on physical constraints rather than spatial interpolation alone.

The ultimate learning objective is to estimate the predictive distribution $p(\mathbf{F}_t \mid \tilde{\mathcal{H}}_{t-1})$ from the partially observed history. Since the clean history $\mathcal{H}_{t-1}$ is unobserved at inference time, we introduce a latent variable view and express the predictive distribution via marginalization over $\mathcal{H}_{t-1}$:
\begin{equation}
\label{eq:predictive_factorization}
p(\mathbf{F}_t \mid \tilde{\mathcal{H}}_{t-1}) = \int p(\mathbf{F}_t \mid \mathcal{H}_{t-1})\; p(\mathcal{H}_{t-1} \mid \tilde{\mathcal{H}}_{t-1}) \;\mathrm{d}\mathcal{H}_{t-1}.
\end{equation}
This decomposition naturally motivates a two-stage learning paradigm. The first term, $p(\mathcal{H}_{t-1} \mid \tilde{\mathcal{H}}_{t-1})$, corresponds to recovering a plausible clean history from corrupted observations, while the second term, $p(\mathbf{F}_t \mid \mathcal{H}_{t-1})$, models the forward wildfire dynamics conditioned on a complete sequence. To improve the computational tractability of the reconstruction phase, we introduce a conditional independence assumption across time. Specifically, we assume that the recovery of the true fire map $\mathbf{F}_\tau$ at any given timestep $\tau$ depends exclusively on the concurrent corrupted observation $\tilde{\mathbf{X}}_\tau^{(\eta)}$, and is conditionally independent of observations at other timesteps.

Since the environmental context $\mathbf{E}_\tau$ is fully observed and provides sufficient instantaneous spatial priors (e.g., topography and vegetation) to guide the inpainting process, the joint reconstruction distribution factorizes over time as:
\begin{equation}
\label{eq:stage1_factorization}
p(\mathcal{H}_{t-1} \mid \tilde{\mathcal{H}}_{t-1}) = \prod_{\tau=t-T}^{t-1} p(\mathbf{F}_\tau \mid \tilde{\mathbf{X}}_\tau^{(\eta)}),
\end{equation}
which yields reconstructed fire maps $\hat{\mathbf{F}}_{t-T:t-1}$. The forecasting stage then learns $p(\mathbf{F}_t \mid \hat{\mathbf{X}}_{t-T:t-1})$, where $\hat{\mathbf{X}}_\tau = [\mathbf{E}_\tau ; \hat{\mathbf{F}}_\tau]$.

\subsection{Stage-I: Fire Shape Reconstruction}
\label{subsec:stage1}

Stage-I aims to reconstruct the latent fire morphology by approximating the conditional distribution $p( \mathcal{H}_{t-1} \mid \tilde{\mathcal{H}}_{t-1})$. Reconstruction is performed independently at each timestep $\tau$, where the model predicts a reconstructed fire map $\hat{\mathbf{F}}_\tau$ from the corrupted state $\tilde{\mathbf{X}}_\tau^{(\eta)}$. To investigate the impact of different inductive biases, we study four representative approaches for finding the reconstruction mapping $f_\theta(\cdot)$, where $\theta$ denotes the set of learnable parameters optimized during the training of the respective neural network architectures.

\begin{itemize}

    \item \textbf{MaskUNet (CNN-based):} Reconstructs missing fire regions by looking primarily at nearby visible patterns and boundaries, and exploiting the resulting spatial locality. Compared with the other methods, it produces a deterministic reconstruction with local factors, but does not explicitly model fire uncertainty or long-range global interactions.
    The MaskUNet reconstruction is formulated as a deterministic mapping:
    \begin{equation}
        \hat{\mathbf{F}}_\tau = f_\theta(\tilde{\mathbf{X}}_\tau^{(\eta)}),
    \end{equation}
    where $f_\theta$ is a convolutional encoder-decoder network. The network $f_\theta$ employs symmetric short cuts between layers (so-called ``skip connections'') to combine low-level, high-resolution spatial features from the encoder directly with the semantically rich, upsampled representations in the decoder~\cite{Ronneberger2015UNet, he2016resnet}. This multi-scale feature aggregation ensures the preservation of fine-grained spatial structure and fire shape details during the inpainting process. 
    \item \textbf{MaskCVAE (Latent Generative):} Captures the uncertainty inherent in stochastic fire boundaries by generating multiple plausible reconstructions from a learned latent representation. To investigate this class of methods, we employ a \ac{cvae}~\cite{kingma2013vae, sohn2015cvae}, which models the complex distribution of possible underlying true fire maps by introducing a continuous latent variable $\mathbf{Z}$. The network is trained by minimizing the negative conditional \ac{elbo}, formulated as the loss objective:
    \begin{equation}
        \begin{aligned}
            \mathcal{L}_{\text{CVAE}} = \;& - \mathbb{E}_{q_\phi(\mathbf{Z} \mid \mathbf{F}_\tau, \tilde{\mathbf{X}}_\tau^{(\eta)})} \left[ \log p_\theta \left(\mathbf{F}_\tau \mid \mathbf{Z}, \tilde{\mathbf{X}}_\tau^{(\eta)}\right) \right] \\
            &+ \beta D_{\text{KL}} \left( q_\phi(\mathbf{Z} \mid \mathbf{F}_\tau, \tilde{\mathbf{X}}_\tau^{(\eta)}) \,\|\, p_\theta(\mathbf{Z} \mid \tilde{\mathbf{X}}_\tau^{(\eta)}) \right),
        \end{aligned}
    \end{equation}
    where $\phi$ and $\theta$ denote the learnable parameters of the recognition and generative models, respectively. Specifically, $q_\phi(\mathbf{Z} \mid \mathbf{F}_\tau, \tilde{\mathbf{X}}_\tau^{(\eta)})$ acts as the probabilistic encoder (approximate posterior) used during training, while $p_\theta(\mathbf{F}_\tau \mid \mathbf{Z}, \tilde{\mathbf{X}}_\tau^{(\eta)})$ serves as the decoder generating reconstructed fire map. The term $p_\theta(\mathbf{Z} \mid \tilde{\mathbf{X}}_\tau^{(\eta)})$ represents the conditional prior network, which allows us to sample latent vectors directly from the corrupted observations during inference. The $D_{\text{KL}}$ operator denotes the \ac{kl} divergence, and $\beta$ is a tunable hyperparameter that balances reconstruction fidelity against the regularization of the latent space. 

    \item \textbf{MaskViT (Transformer-based):} Leverages global context and long-range spatial dependencies by dividing the 2D spatial grid into small non-overlapping patches, flattening each patch into a one-dimensional vector, and treating each vector as a token in a sequence~\cite{vaswani2017attention, dosovitskiy2020vit}. The reconstruction is modeled as:
    \begin{equation}
    \hat{\mathbf{F}}_\tau = g_\theta(\tilde{\mathbf{X}}_\tau^{(\eta)}),
    \end{equation}
    where $g_\theta$ denotes the transformer-based reconstruction network. To dynamically guide the reconstruction of masked regions, $g_\theta$ employs a cross-attention mechanism in which the corrupted fire observations query the fully observed environmental priors. The context-aware fire features are formulated as:
    \begin{equation}
        \text{Attention}(\mathbf{Q}_{\tilde{F}}, \mathbf{K}_E, \mathbf{V}_E) = \operatorname{softmax} \left( \frac{\mathbf{Q}_{\tilde{F}} \mathbf{K}_E^\top}{\sqrt{d_k}} \right) \mathbf{V}_E,
    \end{equation}
    where $\mathbf{Q}_{\tilde{F}}$ represents the query matrix from the tokenized corrupted fire map $\tilde{\mathbf{F}}_\tau^{(\eta)}$, while $\mathbf{K}_E$ and $\mathbf{V}_E$ are the key and value matrices obtained from the multi-modal environmental information $\mathbf{E}_\tau$. The scalar $d_k$ denotes the dimensionality of the key vectors. By relating each fire patch to environmental information across the scene, this mechanism allows the model to draw on unmasked information such as terrain, weather, or vegetation conditions in order to better reconstruct the underlying fire pattern. Distinct from the other models, MaskViT reconstructs missing fire regions by allowing each spatial location to attend to all other locations jointly with the environmental context, thereby enabling scene-wide reasoning over longer ranges.

    \item \textbf{MaskD3PM (Discrete Diffusion):} Frames the fire shape reconstruction as an iterative denoising process operating directly within the discrete state space defined by the binary active/inactive fire pixels~\cite{ho2020denoising, austin2021structured}. The forward corruption process is modeled as a predefined discrete Markov transition matrix with an absorbing state (e.g., a [MASK] token). This mechanism explicitly represents the degradation due to partial observability by progressively transitioning known pixel states into the absorbing mask without cross-flipping the true binary classes. The reverse generative transition at diffusion step $s$ is modeled as a categorical distribution parameterized by a neural network:
    \begin{equation}
    \begin{aligned}
        p_\theta(\mathbf{F}^{(s-1)} \mid \mathbf{F}^{(s)}, \tilde{\mathbf{X}}_\tau^{(\eta)}) = \;& \operatorname{Categorical} \big( \mathbf{F}^{(s-1)}; \\
        & \mathcal{D}_\theta( \mathbf{F}^{(s)}, s, \tilde{\mathbf{X}}_\tau^{(\eta)} ) \big),
    \end{aligned}
    \end{equation}
    where $\mathbf{F}^{(s)}$ denotes the intermediate noisy fire map at the $s$-th diffusion step. The term $\operatorname{Categorical}(\cdot)$ denotes the distribution over the possible discrete pixel classes, applied independently across the $H \times W$ spatial grid. The conditional denoising network $\mathcal{D}_\theta$ takes the current noisy state $\mathbf{F}^{(s)}$, the diffusion timestep $s$, and the corrupted observation $\tilde{\mathbf{X}}_\tau^{(\eta)}$ as inputs to predict the categorical transition logits. By iteratively sampling from this reverse distribution, the model progressively replaces the absorbing masks with inferred physical states to ultimately recover the clean reconstructed fire map, such that $\hat{\mathbf{F}}_\tau = \mathbf{F}^{(0)}$. 
    
\end{itemize}

\subsection{Stage-II: Spatiotemporal Prediction}
\label{subsec:stage2}

Stage-II models the forward dynamics of wildfire spread by learning the predictive distribution $p(\mathbf{F}_t \mid \hat{\mathbf{X}}_{t-T:t-1})$. The input sequence is defined as $\hat{\mathbf{X}}_{t-T:t-1} = \{ \hat{\mathbf{X}}_{t-T}, \dots, \hat{\mathbf{X}}_{t-1} \}$, where each $\hat{\mathbf{X}}_\tau = [\mathbf{E}_\tau ; \hat{\mathbf{F}}_\tau] \in \mathbb{R}^{(C_E+1) \times H \times W}$ integrates the environmental context with the reconstructed fire morphology from Stage-I.
To capture the complex spatiotemporal dependencies, we employ a \ac{utae} framework~\cite{garnot2021panoptic}. The prediction process follows an encoder-decoder structure:

\begin{enumerate}
    \item \textbf{Spatiotemporal Encoding:} The input sequence is processed through a U-Net architecture, in which a multi-scale \ac{cnn} encoder is employed to jointly capture fine-grained local fire morphology and broader scene-level context, both of which are essential for accurately modeling wildfire spread over heterogeneous spatial landscapes. The multi-scale CNN encoder acts as a feature extractor that simultaneously processes the data using convolutional filters of different image resolutions. At each spatial resolution level $l$, corresponding to the progressively downsampled spatial grid $H_l \times W_l$, the encoder generates a sequence of intermediate feature maps $\mathbf{H}^{(l)} = [\mathbf{h}_1^{(l)}, \dots, \mathbf{h}_T^{(l)}]$. Each $\mathbf{h}_k^{(l)}$ represents the spatial distribution of the intermediate feature at the $k$-th timestep. 
    
    To efficiently fuse the temporal sequences while enforcing consistency across spatial scales, a \ac{ltae}~\cite{garnot2020lightweight} is computed only at the bottleneck level, denoted as $L$, corresponding to the deepest encoder stage with the lowest spatial resolution. This level provides the most compact representation of the global scene context. Instead of computing attention independently at every resolution, a single globally learnable master query $\mathbf{Q}$ interacts with the bottleneck keys $\mathbf{K}_k^{(L)}$, derived from the bottleneck feature maps at timestep $k$, to compute a shared temporal attention mask:
    \begin{equation}
    \label{eq:utae_lta_mask}
        \alpha_k = \operatorname{Softmax}\left( \frac{\mathbf{Q} (\mathbf{K}_k^{(L)})^\top}{\sqrt{d_k}} \right),
    \end{equation}
    where $\mathbf{K}_k^{(L)}$ is the key embedding from the $k$-th bottleneck feature map $\mathbf{h}_k^{(L)}$, and $d_k$ is a scaling factor. This bottleneck mask $\alpha_k$ identifies the most informative timesteps from a global semantic perspective. The temporal dimension at any given hierarchy level $l$ is then collapsed into a single summary map $\mathbf{z}^{(l)}$ by progressively upsampling and broadcasting this shared mask:
    \begin{equation}
    \label{eq:utae_lta_fusion}
        \mathbf{z}^{(l)} = \sum_{k=1}^{T} \alpha_k \cdot \mathbf{h}_{k}^{(l)}.
    \end{equation}
    Notably, to maintain a computationally lightweight architecture, an explicit value projection matrix is omitted; instead, the original high-resolution feature map $\mathbf{h}_k^{(l)}$ serves directly as the value ($\mathbf{V}$). By dictating the temporal fusion from the bottleneck, this top-down mechanism ensures that fine-grained spatial details are extracted from the exact same critical moments (e.g., sudden wind shifts) as the global semantics, effectively filtering out residual reconstruction noise from Stage-I.

    \item \textbf{Decoding and Mapping:} The summary maps $\mathbf{z}^{(l)}$ from all levels are fused via skip-connections in a convolutional decoder. The decoder progressively upsamples the features to the original resolution $H \times W$. 
    
    \item \textbf{Final Prediction:} A sigmoid-activated $1 \times 1$ convolution head is applied to the decoder's output to produce the final binary map $\hat{\mathbf{F}}_t \in \{0, 1\}^{H \times W}$, representing the binary active fire for the target day.
\end{enumerate}

This design ensures that the forecasting stage captures local reconstructed fire patterns, environmental constraints, and temporal information.

\section{Dataset and Preprocessing}
\label{sec:dataset}

We evaluate our framework on the \ac{wsts} benchmark dataset to validate its robustness in reconstructing fire morphology and improving forecasting accuracy under partial observability.

\subsection{Dataset Description}
\label{subsec:dataset_details}

The \ac{wsts} dataset~\cite{Gerard2023WildfireSpreadTS} is a large-scale multi-modal benchmark comprising 607 wildfire events across the Western United States (2018--2021). Each sample is a 23-channel spatiotemporal tensor at 375m resolution. As detailed in {Table~\ref{table1}}, the features integrate: (i) \textit{Measurements} (\ac{viirs}~\cite{Vermote2016VNP09GA, Didan2018VNP13A1}, \ac{gridmet}~\cite{Abatzoglou2013GriddedMet}, NASA \ac{srtm}~\cite{NASA2013SRTMGL1}); (ii) \textit{Land Cover} (\ac{modis}~\cite{SullaMenashe2019MCD12Q1C6, FriedlSullaMenashe_MCD12Q1_061_2022}); (iii) \textit{Forecasts} (\ac{gfs}~\cite{Clough2005AER}); and (iv) \textit{Active Fire} maps~\cite{Schroeder2014VIIRS,Oliva2015VIIRSBA}. 

\begin{figure*}[t!]
    \centering
    \includegraphics[width=\linewidth]{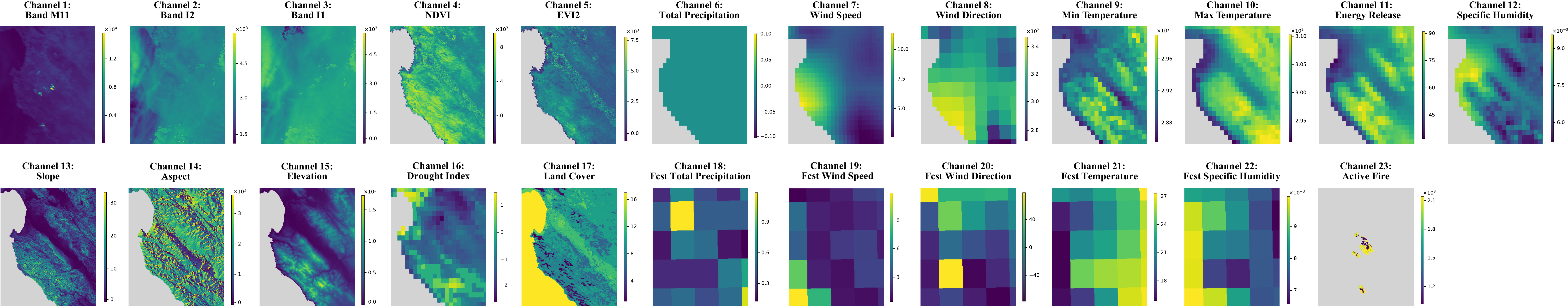}
    \caption{\small{Visualization of the 23-channel input features for a single daily observation of the Dolan Fire event in Monterey County, California, from Aug 14 to Aug 31, 2020. Grey regions indicate missing or undetected pixels (NaNs). Each subplot employs an independent color scale, identical colors across panels may represent different physical magnitudes.}}
    \label{fig:23_channels}
\end{figure*}

\begin{figure}[t!]
    \centering
    \includegraphics[width=\linewidth]{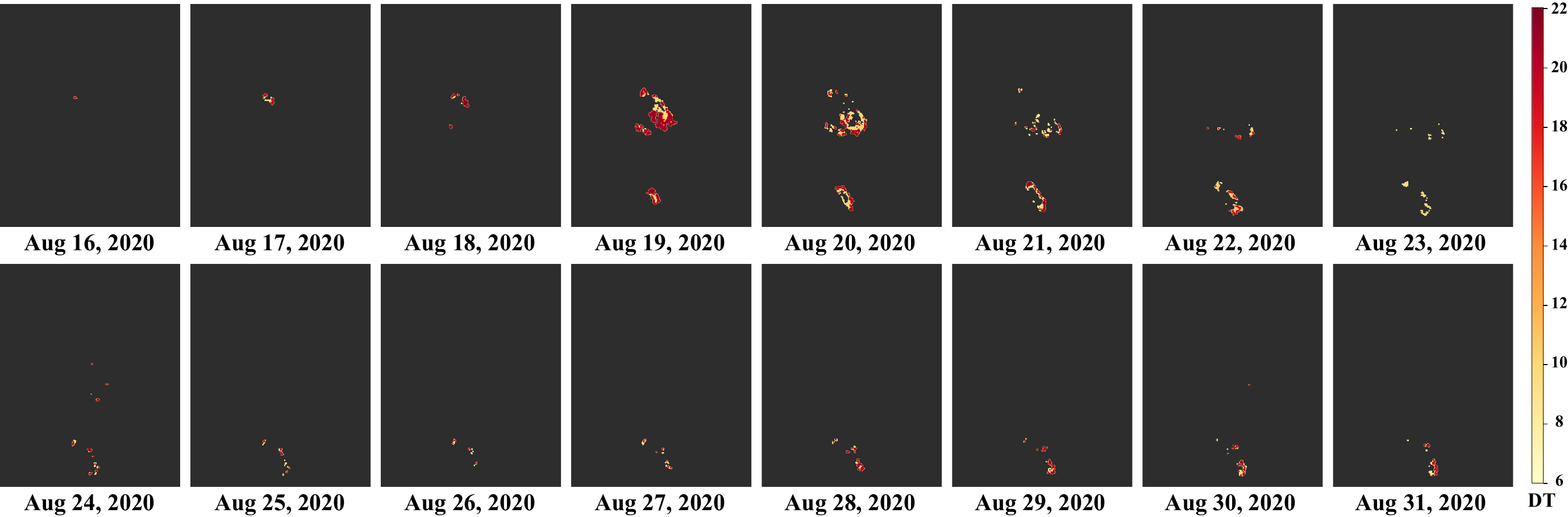}
    \caption{\small{Daily active fire detections for the Dolan Fire (Aug 16--31, 2020), where colored pixels indicate detection time and black denotes missing or undetected (NaN).}}
    \label{fig:fire_visualization}
\end{figure}

Figs.~\ref{fig:23_channels} and~\ref{fig:fire_visualization} provide two complementary perspectives on the Dolan Fire in Monterey County, California. Fig.~\ref{fig:23_channels} displays all 23 input channels for a single daily observation. All channels are gridded at a uniform spatial resolution of 375 m. Fig.~\ref{fig:fire_visualization} complements this static snapshot by depicting the day-by-day progression of active fire detections from August 16--31, 2020. Across the 16-day sequence, active fire pixels remain confined to a small fraction of the spatial window, illustrating the intrinsic sparsity of satellite-based fire observations and the severe class imbalance that any predictive model must address.


\subsection{Preprocessing and Feature Engineering}
\input{tables/table1}
To mitigate extreme class imbalance and strictly focus the learning process on the evolving fire front, we employ a content-adaptive cropping strategy. The variable-sized satellite imagery is converted into fixed-size spatiotemporal sequences, defined by a spatial window of $64 \times 64$ pixels and a temporal window of $T=5$ days. To determine the optimal crop location, we implement a temporally-prioritized search mechanism for each training pair. We iterate backward through the historical window and compute the spatial centroid of the active fire pixels from the most recent day with valid fire detections. This centroid serves as the sequence center. If no active fire is present across the entire 5-day history, the crop defaults to the geometric center of the original image tile. Both the input sequence and the target label are extracted using these identical coordinates to guarantee strict spatiotemporal alignment.

\input{tables/table2}
The original 23 raw sensor channels are transformed into a 42-channel feature tensor through three explicit preprocessing steps, designed to convert heterogeneous physical variables into a unified model-ready representation. Starting from 23 raw channels, the final 42-channel tensor is obtained by expanding the two angular channels into four cyclical channels in total, converting the single land-cover channel into 17 one-hot channels, and retaining the remaining 21 channels as normalized continuous features. The resulting transformations are summarized as follows:
\begin{itemize}
    \item \textbf{Cyclical Encoding:} The two angular variables, Wind Direction and Aspect, are each transformed from a single scalar channel into a two-channel $[\sin(\theta), \cos(\theta)]$ representation to preserve periodic continuity and avoid artificial discontinuities at the angular wraparound.
    \item \textbf{Categorical Expansion:} Land Cover (originally Channel 17) is converted from a single categorical channel into a 17-channel one-hot representation, so that each land-cover type is represented independently without introducing spurious ordinal relationships.
    \item \textbf{Statistical Normalization:} All remaining continuous channels retain their original dimensionality and are standardized using Z-score normalization based on the dataset statistics reported in Table~\ref{table1}. Any NaN values are set to zero after normalization.
\end{itemize}

\subsection{Evaluation Scenarios}
\begin{figure}[t!]
    \centering
    \includegraphics[width=\columnwidth]{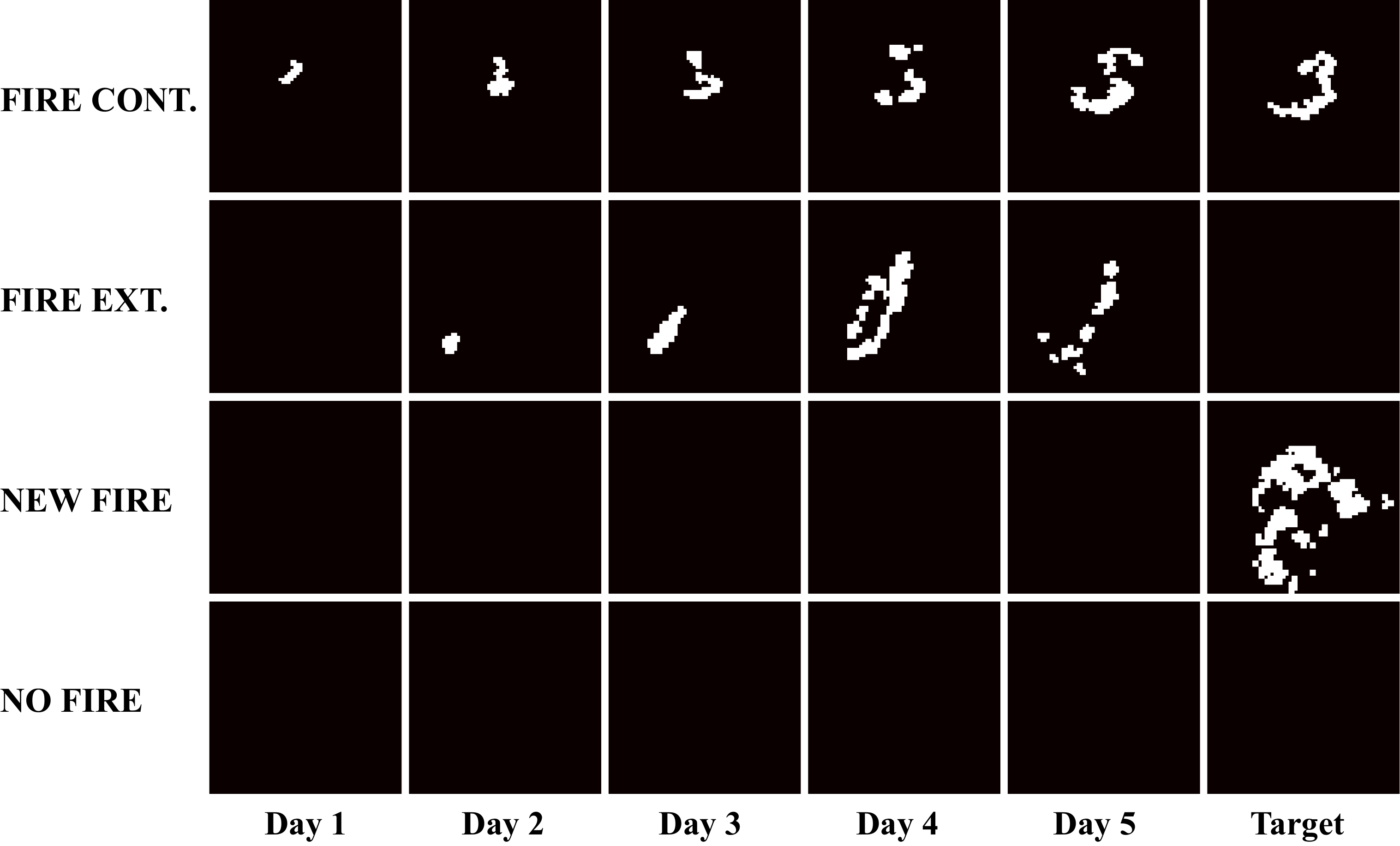}
    \caption{
    \textbf{Definition of the four evaluation scenarios in \ac{wsts}.} Each row shows a binary 5-day historical fire sequence together with the prediction target, where white regions denote active fire pixels.}
    \label{fig:fire_scenarios}
\end{figure}

Following the \ac{wsts} protocol, each sample is categorized into one of four fire-dynamics states to reflect different evolution patterns: \textsc{Fire Continues}, \textsc{Fire Extinguished}, \textsc{New Fire with No History}, and \textsc{No Fire}. Specifically, \textsc{Fire Continues} denotes sequences where active fire persists into the prediction day, while \textsc{Fire Extinguished} corresponds to cases where a previously observed fire disappears. \textsc{New Fire with No History} captures ignition events absent from the historical window, and \textsc{No Fire} represents background samples with no fire activity. These scenario splits allow separate evaluation of reconstruction fidelity and over-prediction behavior under partial observability. A representative visualization of the four scenarios is shown in {Fig.~\ref{fig:fire_scenarios}}.

\subsection{Partial Observability Simulation}
We simulate sensor degradation by applying a corruption operator $\mathcal{C}_\eta$ that masks a fraction $\eta \in \{0.1, 0.2, \ldots, 0.8\}$ of fire-channel pixels, where $\eta$ denotes the ratio of masked fire pixels to total fire pixels. Two masking mechanisms are considered. In \textit{pixel-wise masking}, each fire pixel is independently assumed to be blocked with probability $\eta$, producing spatially uncorrelated gaps that simulate random sensor noise. In \textit{block-wise masking}, contiguous rectangular patches of size $s \times s$, where $s$ is an integer in $[4, 16]$, are iteratively removed from fire-active regions until the cumulative masked fraction reaches $\eta$, simulating spatially coherent occlusions such as cloud cover or smoke obstruction.

\section{Experimental Results}
\label{sec:experiments}

\subsection{Experiment Setup}

Our experimental evaluation is divided into two stages: Stage-I for observation reconstruction and Stage-II for wildfire spread prediction. We evaluate the performance under two distinct types of data corruption: pixel-wise and block-wise occlusions. Furthermore, we assess the robustness of the models across varying degradation levels, with a missing data ratio $\eta$ ranging from 10\% to 80\%. Our primary focus in the experiments is to demonstrate the substantial performance improvement achieved by predicting on data fully reconstructed by Stage-I, as opposed to directly forecasting from the corrupted observations. The complete configuration and training details can be found in the Appendix. A Pytorch implementation is available at~\cite{githubrepo}.

For the observation reconstruction task, we evaluate a total of six different models, comprising four learning-based methods and two non-learning baselines. As detailed previously, the four learning-based models span diverse generative paradigms: a Residual U-Net (MaskUNet), a Conditional VAE (MaskCVAE), a Vision Transformer (MaskViT), and a discrete diffusion model (MaskD3PM). To benchmark these approaches, we introduce two non-learning baselines, which are implemented as follows:
\begin{itemize}
    \item \textbf{Random}: Masked pixels are filled with independent Bernoulli samples ($p{=}0.5$), representing a no-information lower bound.
    \item \textbf{Dilation}: A morphological dilation operator is applied to the currently observed fire pixels using a kernel with a radius of 5, propagating the nearest visible signal into the occluded regions.
\end{itemize}

We evaluate the reconstruction performance across four distinct fire evolution scenarios (Fig.~\ref{fig:fire_scenarios}), employing specific metrics tailored to the characteristics of each scenario. To prevent artificial inflation from uncorrupted pixels, all reconstruction metrics are calculated exclusively within the occluded region~$\mathcal{O}$:
\begin{itemize}
    \item \textbf{\ac{fpr}}: For the \textsc{No Fire} and \textsc{New Fire} scenarios, there are no active fires in the historical sequence. Our primary concern is to strictly avoid generating spurious fires from nothing. Therefore, we use \ac{fpr} to quantify over-prediction, defined as:
    \begin{equation}\label{eq:fpr}
        \text{FPR} = \frac{\text{FP}_{\mathcal{O}}}{\text{FP}_{\mathcal{O}} + \text{TN}_{\mathcal{O}}},
    \end{equation}
    where $\text{FP}_{\mathcal{O}}$ and $\text{TN}_{\mathcal{O}}$ denote the number of false positives and true negatives, respectively, calculated strictly within the mask $\mathcal{O}$.
    \item \textbf{Dice Coefficient}: For the \textsc{Fire Continues} and \textsc{Fire Extinguished} scenarios, the historical observations contain a significant number of fire pixels. We use the Dice coefficient, which is a specialized metric for evaluating morphological overlap in highly imbalanced spatial data. It is defined as:
    \begin{equation}\label{eq:dice}
        \text{Dice} = \frac{2\,|\hat{\mathbf{F}}_{\mathcal{O}} \cap \mathbf{F}_{\mathcal{O}}|}{|\hat{\mathbf{F}}_{\mathcal{O}}| + |\mathbf{F}_{\mathcal{O}}|},
    \end{equation}
    where $\hat{\mathbf{F}}_{\mathcal{O}}$ and $\mathbf{F}_{\mathcal{O}}$ denote the predicted and ground-truth fire pixels within $\mathcal{O}$, respectively.
\end{itemize}

In the prediction stage, we employ \ac{utae} as our forecasting backbone. Similarly, we test the prediction performance across the same four scenarios. However, the evaluation metrics are adapted based on the presence of fire on the final target day, as discussed below.
\begin{itemize}
    \item \textbf{\ac{ap}}: For the \textsc{Fire Continues} and \textsc{New Fire} scenarios, active fires exist on the target day. We evaluate the models using \ac{ap}, which computes the area under the Precision-Recall (PR) curve. We use \ac{ap} instead of the Dice coefficient for Stage-II because the forecasting model outputs continuous probability maps rather than deterministic binary masks. While Dice evaluates the discrete quality of a single thresholded recovery, \ac{ap} assesses the overall ranking and confidence distribution of the forecasting probabilities, providing a more comprehensive measure of the model's overall predictive capability across different confidence levels.
    \item \textbf{\ac{fpr}}: For the \textsc{No Fire} and \textsc{Fire Extinguished} scenarios, since the target day contains zero fire pixels, we continue to use \ac{fpr} in~\eqref{eq:fpr} to penalize any false alarms generated by the prediction model.
\end{itemize}

Having established the experimental configuration and evaluation metrics, we now present a comprehensive analysis of our proposed two-stage framework. The subsequent sections systematically dissect the pipeline's performance across varying models, physical scenarios, and degradation severities. Specifically, in \textbf{Section \ref{sec:exp_stage1}}, we focus on the Stage-I observation recovery, comparing the structural reconstruction capabilities of the different generative networks against the non-learning baselines. This section details how different inductive biases handle localized (pixel-wise) versus contiguous (block-wise) missing data across the four distinct fire evolution scenarios. Following this, \textbf{Section \ref{sec:exp_stage2}} investigates the ultimate goal of our framework: Stage-II wildfire forecasting. We quantify the critical impact of the preceding recovery phase by explicitly comparing downstream predictive performance on raw corrupted inputs versus Stage-I reconstructed sequences.

\subsection{Stage-1: Reconstruction Performance}
\label{sec:exp_stage1}
\input{tables/table3}

We evaluate the Stage-I recovery models across four distinct physical scenarios under both pixel-wise and block-wise masking mechanisms. It is crucial to note that for Stage-I historical recovery, both the \textsc{Fire Continues} and \textsc{Fire Extinguished} scenarios evaluate the reconstruction of currently active fire pixels. They are partitioned based on their future role in assessing how well the models preserve the advancing fire fronts versus their decaying trailing edges. The quantitative results for these active regions are presented in Table~\ref{tab:stage1}. Conversely, the \textsc{New Fire} and \textsc{No Fire} scenarios are analyzed textually due to their near-zero error bounds among the learning-based approaches. Because the historical sequences in these scenarios are entirely devoid of active fires, the models observe zero fire in the unoccluded regions, regardless of the masking severity. Consequently, the learning-based algorithms reliably infer a almost completely fire-free state without hallucinating non-existent fires, strictly suppressing false alarms in unburned backgrounds.

\subsubsection{Reconstruction of Advancing Fire Fronts (\textsc{Fire Continues})} 
The \textsc{Fire Continues} subset represents the active, propagating perimeters of the fire. Reconstructing these highly complex and fragmented morphological structures is inherently challenging. We analyze the performance in the following ways:
\begin{itemize}
    \item \textbf{Analysis of Pixel-wise Masking:} Localized context is paramount. Both MaskCVAE and MaskUNet demonstrate superiority. At a low masking ratio ($\eta=10\%$), they achieve Dice scores of $0.924$ and $0.920$, respectively, vastly outperforming the deterministic Dilation baseline ($0.410$). As the image corruption scales to the extreme value $\eta=80\%$, MaskCVAE maintains a robust Dice of $0.747$, proving that its learned latent prior provides stable constraints for complex front morphology. Conversely, the discrete diffusion model MaskD3PM degrades significantly to $0.376$, struggling to maintain spatial coherence under extreme sparsity.
    \item \textbf{Analysis of Block-wise Masking:} Large-scale occlusions severely penalize purely local models when the advancing front is entirely masked. While MaskCVAE ($0.598$ at $\eta=80\%$) remains the leading approach, MaskViT exhibits remarkable resilience. Despite beginning with worse performance at $\eta=10\%$ masking, MaskViT yields a competitive Dice of $0.488$ at $80\%$ masking. This highlights the advantage of its cross-attention mechanism, which can attend to distant environmental cues to infer the missing front.
\end{itemize}
\subsubsection{Reconstruction of Decaying Trailing Edges (\textsc{Fire Extinguished})} 
The \textsc{Fire Extinguished} subset represents currently burning areas that will soon burn out, typically corresponding to the contiguous, internal smoldering cores of the fire. Because these regions are more spatially continuous than advancing fronts, they yield inherently higher Dice scores across all models. We analyze the performance in the following ways:
\begin{itemize}
    \item \textbf{Analysis of Pixel-wise Masking:} The MaskCVAE and MaskUNet models achieve near-perfect recovery ($\sim 0.95$ Dice) at $\eta=10\%$. Even under severe $\eta=80\%$ occlusion, both models sustain high fidelity ($\sim 0.79$ Dice), indicating that dense, localized spatial context is highly effective at filling in missing internal fire pixels.
    \item \textbf{Analysis of Block-wise Masking:} The performance degradation is more graceful here than in the \textsc{Fire Continues} scenario. MaskCVAE continues to outperform the other methods ($0.650$ at $\eta=80\%$), while MaskUNet, MaskViT, and MaskD3PM all cluster closely between $0.57$ and $0.65$ under extreme block occlusion. This suggests that, unlike the complex advancing fronts, reconstructing contiguous trailing edges is generally less sensitive to the specific architectural inductive biases of the models, whether it be the localized operations of \acp{cnn} or the global attention of Transformers.
\end{itemize}
\subsubsection{Specificity and False Alarm Suppression (\textsc{New Fire} \& \textsc{No Fire})}
To optimize layout and focus on complex propagation dynamics, tabular results for the \textsc{New Fire} and \textsc{No Fire} scenarios are omitted from the main text. These scenarios predominantly consist of unburned background regions, strictly testing the models' resistance to hallucination.

Under both pixel-wise and block-wise masking, the random baseline yields an expected \ac{fpr} of approximately $0.500$, while the Dilation baseline provides more reasonable performance. In contrast, all of the evaluated learning structures (MaskCVAE, MaskUNet, MaskViT, MaskD3PM) demonstrate exceptional specificity. Their \acp{fpr} remain strictly bounded near zero (consistently below $0.001$) across all masking levels $\eta$. This confirms that our Stage-I recovery models robustly suppress false alarms and strictly avoid hallucinating non-existent fire pixels in unburned areas, ensuring that no ``phantom fires" are passed to the Stage-II forecasting module.
\subsubsection{Model Comparison}
\begin{figure*}[!t]
\centering
\begin{minipage}{0.48\textwidth}
    \centering
    \includegraphics[width=\linewidth]{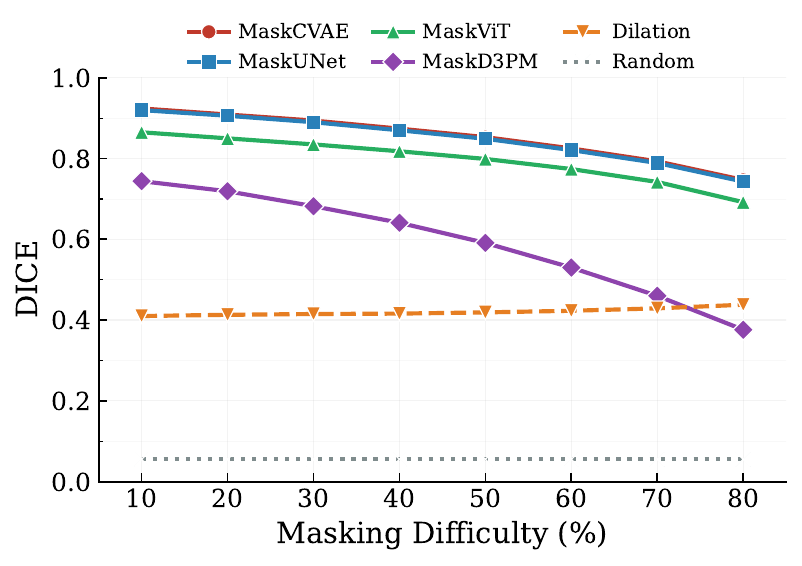}
    \vspace{-0.5cm} 
    
    {\small \hspace{0.6cm} \textbf{(a) PIXEL-WISE RECOVERY}}
    \label{fig:recover_model_compare_pixel}
\end{minipage}
\hfill 
\begin{minipage}{0.48\textwidth}
    \centering
    \includegraphics[width=\linewidth]{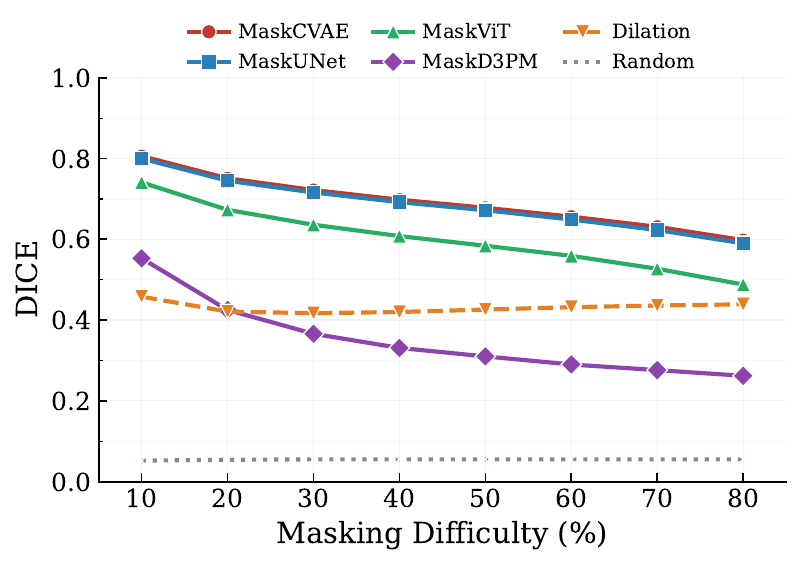}
    \vspace{-0.5cm} 
    
    {\small \hspace{0.75cm} \textbf{(b) BLOCK-WISE RECOVERY}}
    \label{fig:recover_model_compare_block}
\end{minipage}
\caption{Comparison of Stage-I recovery performance across varying masking difficulty levels ($\eta \in [10\%, 80\%]$). The performance is evaluated using the \textbf{DICE coefficient}, which is particularly effective for capturing the structural integrity of \textbf{sparse flame patterns}. In both (a) and (b), \textit{Random} and \textit{Dilation} represent the \textbf{baseline methods}, whereas the other four curves denote different \textbf{learning-based models}. The results highlight the superior recovery performance of learning-based approaches under high-sparsity conditions.}
\label{fig:stage1_curves}
\end{figure*}

To ensure a rigorous evaluation of spatial reconstruction, we strictly isolate the analysis to samples containing active fire pixels. Calculating the Dice coefficient on completely unburned images can lead to numerical instability and artificially skew the overall metric. By filtering out these empty samples, we guarantee that the reported Dice scores faithfully reflect the models' ability to reconstruct the complex fire behavior. The overall reconstruction performance under this refined criterion across all learning networks is illustrated in {Fig. \ref{fig:stage1_curves}}.

\subsection{Stage-2: The Impact of Recovery on Forecasting}
\label{sec:exp_stage2}
\input{tables/table5}

Table~\ref{tab:stage2} provides a scenario-wise breakdown of downstream forecasting performance under both pixel-wise and block-wise corruptions. To maintain the focus of our analysis, we exclusively tabulate the \textsc{Fire Continues} and \textsc{Fire Extinguished} scenarios. The \textsc{New Fire} and \textsc{No Fire} scenarios are omitted since their historical input sequences contain entirely unburned regions; applying spatial masks to these empty historical inputs yields no data alteration, resulting in identical prediction performance across all masking severity levels. Thus, Stage-I reconstruction provides no variable advantage when a historical fire is absent. For the \textsc{New Fire} scenario, the \ac{ap} remains constant at 0.096$\pm${0.152} (pixel-wise) and 0.123$\pm${0.183} (block-wise), reflecting that predictions in these areas rely entirely on environmental ignition cues rather than reconstructed morphology. For the \textsc{No Fire} scenario, the model remains highly robust, maintaining a near-zero \ac{fpr} consistently below 0.001 across all settings.

Focusing on the tabulated scenarios containing active historical fire, the forecasting model exhibits distinct sensitivities to data corruption. For \textsc{Fire Continues}, the \ac{ap} decreases monotonically as masking severity increases. It drops from an initial value of 0.527 without occlusion to 0.385 under 80\% pixel-wise masking, and suffers a steeper decline to 0.334 under 80\% block-wise masking. This substantial degradation highlights that accurate downstream forecasting is highly dependent on the morphological integrity of the advancing fire fronts. Conversely, in the \textsc{Fire Extinguished} scenario, the \ac{fpr} remains near zero. This indicates that when burnout patterns dominate the historical sequence, the model is remarkably resilient to information loss and rarely hallucinates subsequent fire spread.

\begin{figure}[t]
    \centering
\includegraphics[width=\linewidth]{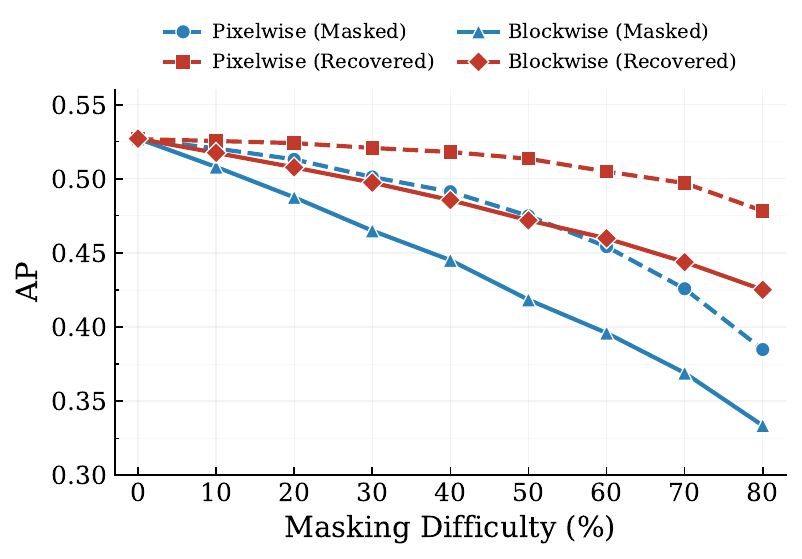}
    \caption{
    Forecasting performance under increasing masking difficulty. Average Precision (AP) is reported for direct forecasting on corrupted inputs (Masked)  nd forecasting after Stage-I recovery (Recovered) under both pixel-wise and block-wise corruption. Recovery consistently reduces performance degradation, especially under severe structural occlusion.
    }
    \label{fig:stage2_curves}
\end{figure}

Fig.~\ref{fig:stage2_curves} demonstrates the substantial improvement in next-day forecasting performance achieved by integrating the Stage-I reconstruction module across varying masking conditions. Compared to direct prediction on corrupted inputs, forecasting on recovered sequences yields significant and consistent performance gains, particularly under severe data loss. Specifically, under extreme 80\% pixel-wise masking, the forecasting \ac{ap} on raw corrupted data plummets to 0.378. In contrast, predicting on the reconstructed sequences restores the \ac{ap} to 0.482, representing a substantial relative improvement of 27.5\%. Similarly, under 80\% block-wise occlusion, where spatially contiguous data loss more severely disrupts image integrity, the unrecovered prediction AP drops to 0.328. Following reconstruction, the forecasting performance recovers to 0.425, achieving a substantial relative increase of 29.6\%. These quantitative gains confirm that restoring missing observations prior to forecasting drastically mitigates the domain gap caused by partial observability.

\begin{figure*}[t!]
    \centering
    \includegraphics[width=\linewidth]{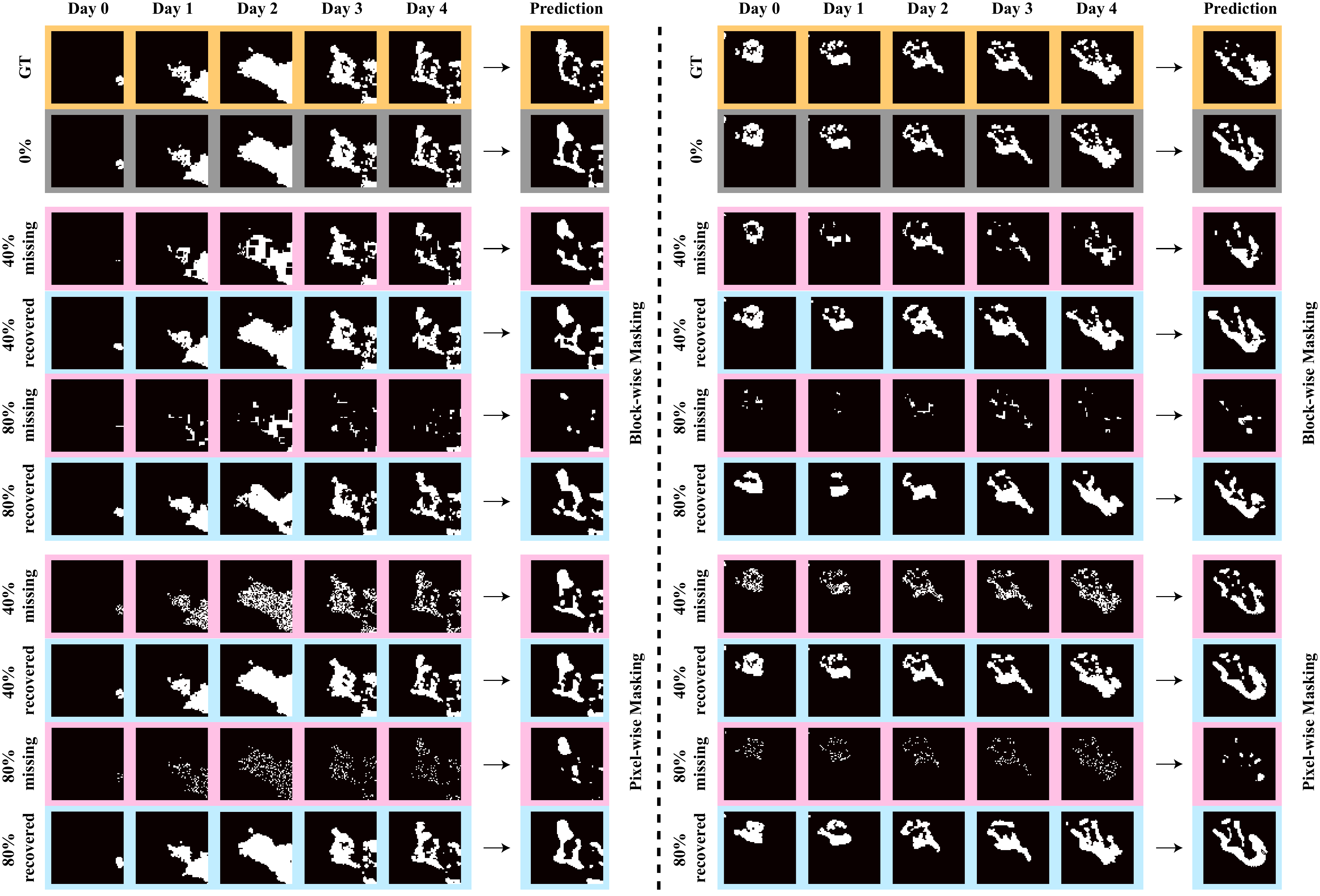}
    \caption{\small{\textbf{Visualization of the MaskUNet recovery-and-prediction pipeline under different information loss (Pixel-wise vs. Block-wise).}
    The sequence illustrates the wildfire forecasting task for the \textbf{next day} based on historical observations from the preceding \textbf{five days}. To clarify the visual layout: \textbf{orange} borders denote the actual Ground Truth (\textbf{GT}) fire footprint on the target day, representing the real-world physical outcome. In contrast, the \textbf{gray} borders (labeled as \textbf{0\%}) represent the model's prediction when provided with completely uncorrupted historical observations. Furthermore, \textbf{pink} borders highlight the (\textbf{Missing}) input after masking, while \textbf{blue} borders indicate the fire shape details reconstructed by our model (\textbf{Recovered}) and the final output for the next-day state (\textbf{Prediction}).}}
    \label{fig:visualization}
\end{figure*}

Fig.~\ref{fig:visualization} illustrates the reconstruction capability of MaskUNet pipeline through two representative examples under pixel-wise and block-wise occlusion. In scenarios where the fire front is severely obscured, the recovered maps successfully capture the fire's shape by synthesizing sparse detections with contextual information. By restoring the spatial connectivity of the fire front, these recovered features provide a highly robust initial state for downstream forecasting. {Fig.~\ref{fig:visualization}} contrasts our pipeline's results against predictions made using completely uncorrupted historical data, which serve as the performance upper bound for our forecasting module. As demonstrated in both examples, our learning-based recovery-and-prediction approach leads to superior alignment with the Ground Truth (GT). Compared to direct forecasting from missing data, which severely underestimates fire spread, our approach bridges the gap caused by information loss and successfully pushes the prediction performance closer to the performance upper bound.

\section{Conclusion}
\label{sec:conclusion}

This paper has addressed the critical but under-explored challenge of wildfire spread forecasting under partial observability. To mitigate the domain gap caused by incomplete satellite observations, we proposed a two-stage framework that decouples observation recovery from spatiotemporal prediction. By systematically evaluating diverse generative algorithms for the Stage-I reconstruction, we demonstrated that learning-based inpainting models, particularly those leveraging strong local inductive biases and latent priors, can effectively restore complex fire features even under severe spatial occlusion. Notably, our Stage-II forecasting evaluations confirmed that integrating this dedicated recovery module significantly reduces the performance degradation inherent in corrupted inputs, successfully pushing the downstream predictive accuracy back toward the upper performance bound achievable with uncorrupted data. 

While our decoupled approach significantly enhances operational forecasting reliability, several avenues remain for future exploration. Subsequent research should incorporate physically realistic image corruption models derived from actual cloud and smoke transport dynamics to better reflect real-world sensor obscuration. Additionally, transitioning from independent stage training to an end-to-end joint optimization could enable the recovery module to directly maximize downstream forecasting accuracy rather than strictly pixel-level fidelity. Finally, extending this framework to accommodate higher-resolution satellite products and extended prediction horizons will further broaden its utility for real-time wildfire management systems.

\appendix
\section{Implementation Details}
\label{app:implementation}

This appendix specifies the exact architectural configurations and optimization details required to reproduce the experiments. 
High-level modeling principles are described in Section~\ref{sec:method}.
The code implementation can be found at~\cite{githubrepo}.

\subsection{Common Architectural Settings}

Unless otherwise stated, all convolutional layers use:

\begin{itemize}
\item Kernel size $3\times3$ with padding 1
\item Group Normalization (8 groups)
\item SiLU activation
\item $1\times1$ convolution for output logits
\end{itemize}
All models operate on $64\times64$ spatial inputs with a $42$-channel tensor after preprocessing.

\subsection{Stage-I Reconstruction Backbones}

\subsubsection{MaskUNet}

\begin{itemize}
\item Input: $42\times64\times64$
\item Encoder channels: $[64,128,256,512]$
\item Residual blocks per scale: 2
\item Downsampling: $64\rightarrow32\rightarrow16\rightarrow8\rightarrow4$
\item Bottleneck resolution: $4\times4$
\item Multi-head self-attention at bottleneck
\item Decoder: symmetric upsampling with skip connections
\item Output: $1\times64\times64$ logits
\end{itemize}

\subsubsection{MaskCVAE}

\begin{itemize}
\item Encoder channels: $[64,128,256]$
\item Spatial resolution: $64\rightarrow32\rightarrow16\rightarrow8$
\item Latent dimension: 128
\item Gaussian prior and posterior parameterization $(\mu,\log\sigma)$
\item Latent fusion via channel concatenation + residual block
\item Decoder: UNet-style upsampling
\item Bottleneck resolution: $8\times8$
\item KL weight: $\beta=0.1$
\end{itemize}

\subsubsection{MaskViT}

\begin{itemize}
\item Patch size: $4\times4$
\item Token grid: $16\times16$ (256 tokens)
\item Embedding dimension: 256
\item Transformer depth: 6
\item Attention heads: 8
\item MLP expansion ratio: 4
\item Pre-LayerNorm configuration
\item Linear projection head + spatial reshape
\end{itemize}

\subsubsection{MaskD3PM}

\begin{itemize}
\item Diffusion steps: 100
\item State space: $\{0,1,\text{MASK}\}$
\item Sinusoidal time embedding dimension: 256
\item Backbone: ResUNet with AdaGN conditioning
\item Bottleneck resolution: $8\times8$
\item Reverse transition modeled with categorical logits
\end{itemize}

\subsection{Stage-II Spatiotemporal Predictor}

The Stage-II forecasting model follows a U-TAE style encoder–decoder design.

\begin{itemize}
\item Input sequence length: $T=5$
\item Encoder channels: $[64,128,128]$
\item Downsampling: $64\rightarrow32\rightarrow16\rightarrow8$
\item Temporal attention heads: 4
\item Dropout: 0.1
\item Decoder: skip-connected upsampling
\item Output: sigmoid-activated $1\times1$ convolution
\end{itemize}

\subsection{Optimization and Loss Functions}

All models are trained using the same optimization protocol unless otherwise specified.

\begin{itemize}
\item Optimizer: AdamW
\item Learning rate: $1\times10^{-4}$
\item Weight decay: $1\times10^{-2}$
\item Batch size: 16
\item Training epochs: 100
\item Early stopping patience: 10 epochs
\item Learning rate schedule: cosine decay
\item Mixed precision: enabled
\item Hardware: NVIDIA A100 GPU
\end{itemize}

Loss functions are defined as follows:

\begin{itemize}
\item MaskUNet / MaskViT: Focal loss
\item MaskCVAE: ELBO (Focal reconstruction + KL term)
\item MaskD3PM: Categorical cross-entropy
\item Stage-II forecasting: Focal loss
\end{itemize}
\subsection{Corruption and Masking Implementation}

For reconstruction robustness experiments, input corruption is applied before Stage-I training:

\begin{itemize}
\item Masking ratios: $\eta\in\{0.1,0.2,\dots,0.8\}$
\item Pixel-wise corruption: independent Bernoulli masking
\item Block-wise corruption: contiguous square regions sampled with weights
\item Corrupted values replaced with a dedicated [MASK] token
\end{itemize}



\bibliographystyle{IEEEtran}
\bibliography{ref}

\begin{IEEEbiography}[{\includegraphics[width=1.25in,height=1.25in,clip,keepaspectratio]{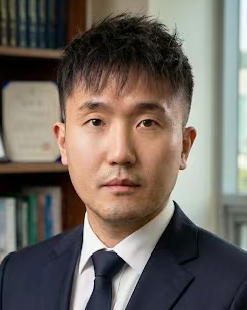}}]
{Chen Yang} received the B.S. degree in Communication Engineering from Beijing University of Posts and Telecommunications, Beijing, China, in 2017. He is currently pursuing the Ph.D. degree in Electrical Engineering and Computer Science with the University of California, Irvine, CA, USA. His current research focuses on data assimilation, Bayesian inference/estimation, generative models, and autonomous UAV sensing.
\end{IEEEbiography}

\begin{IEEEbiography}[{\includegraphics[width=1.25in,height=1.25in,clip,keepaspectratio]{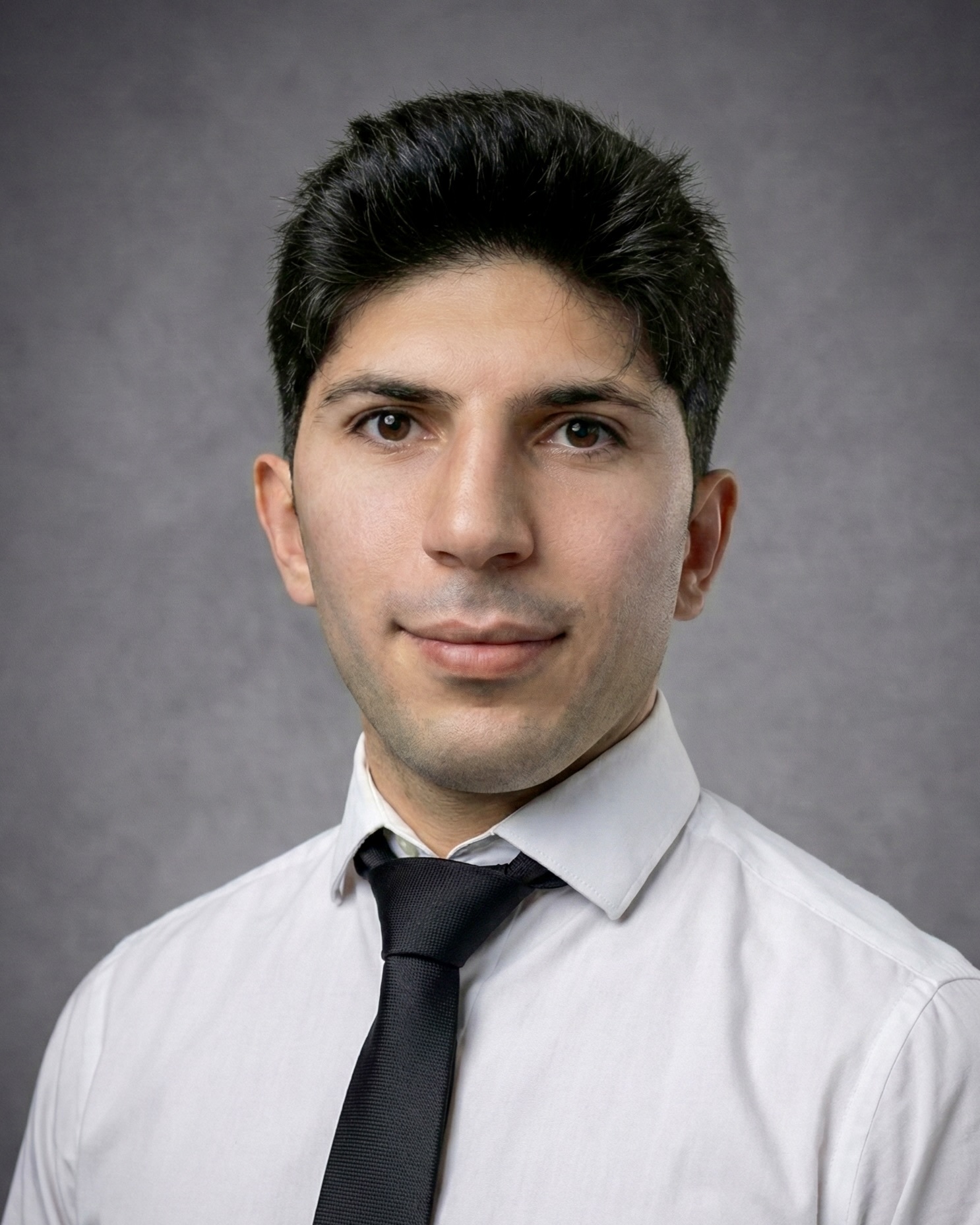}}]
{Mehdi Zafari} (S'21) received the B.S. degree in Electrical Engineering from the University of Tehran, Tehran, Iran, in 2015, and the M.S. degree in Electrical and Computer Engineering from Rice University, Houston, TX, USA, in 2024. He is currently pursuing the Ph.D. degree in Electrical Engineering and Computer Science at the University of California, Irvine, CA, USA.
His research interests include optimization, deep learning, and generative AI for next-generation wireless communications, with a focus on integrated sensing and communication (ISAC), cell-free/distributed MIMO technology, and resource allocation.
\end{IEEEbiography}

\begin{IEEEbiography}[{\includegraphics[width=1.25in,height=1.25in,clip,keepaspectratio]{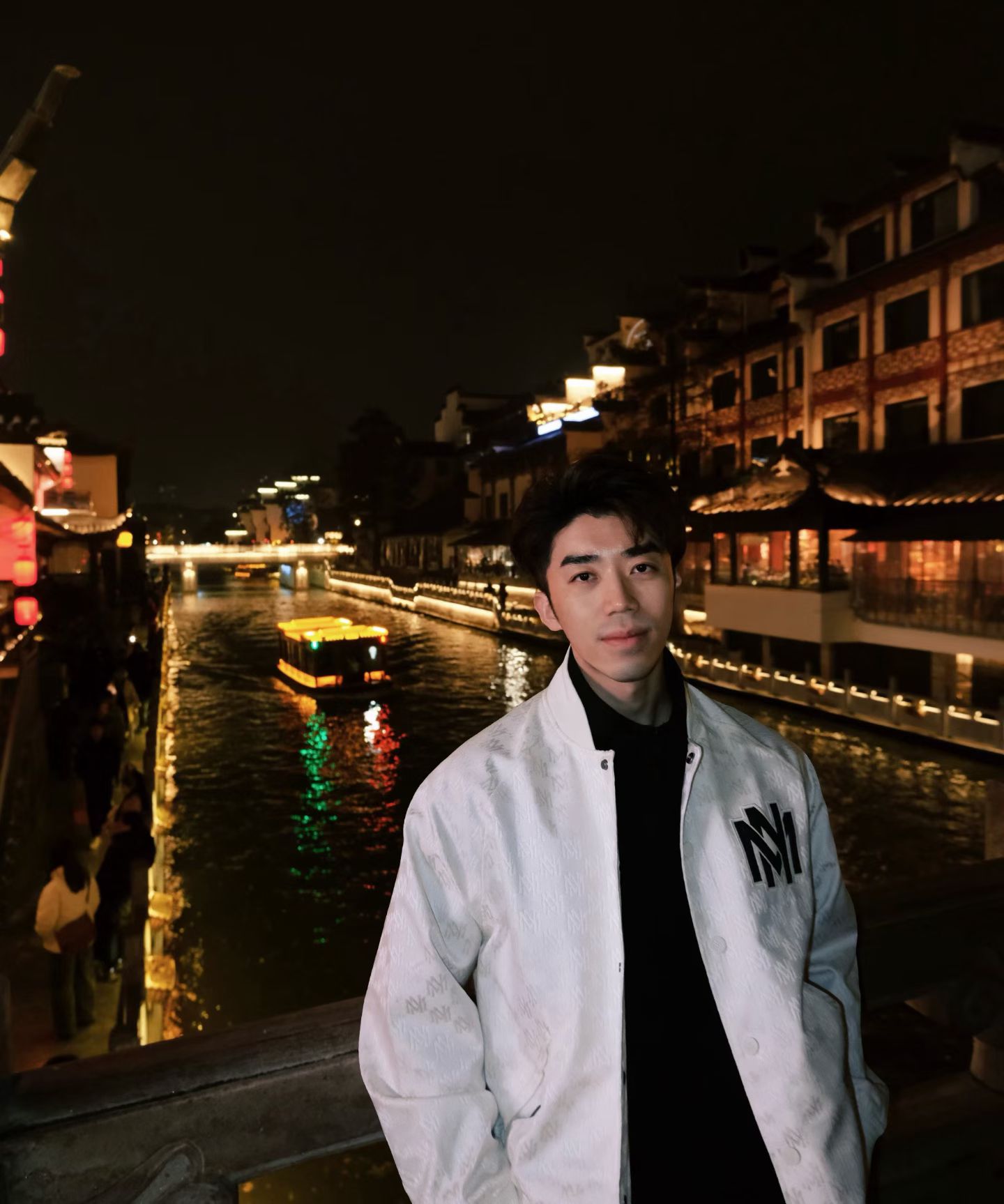}}]
{Ziheng Duan} received the Ph.D. degree in Computer Science from the University of California, Irvine. His research interests include machine learning and its applications to real-world problems.
\end{IEEEbiography}

\begin{IEEEbiography}[{\includegraphics[height=1.25in,clip,keepaspectratio]{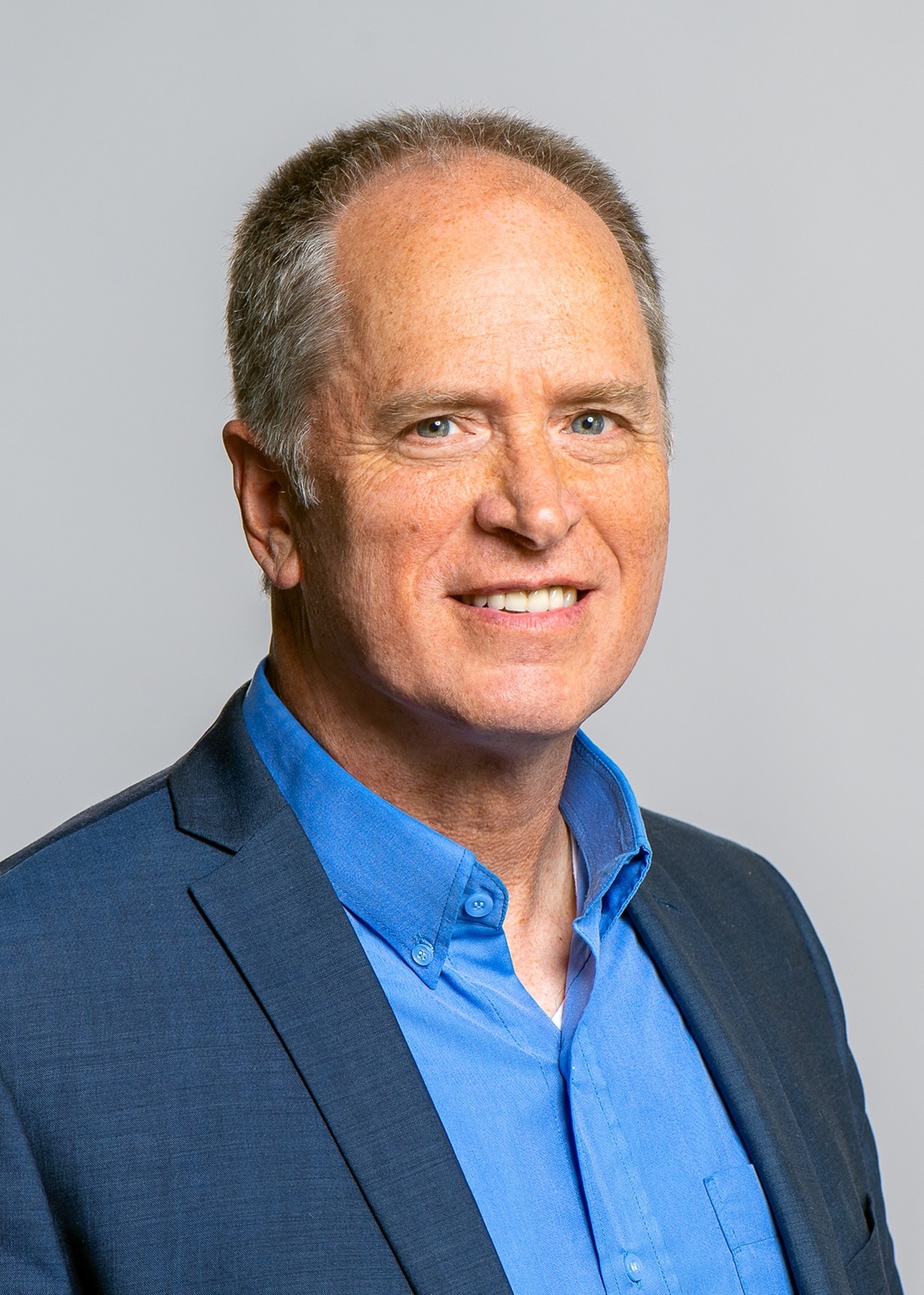}}]
{A. Lee Swindlehurst} (Life Fellow, IEEE) received the B.S. (1985) and M.S. (1986) degrees in Electrical Engineering from Brigham Young University (BYU), and the PhD (1991) degree in Electrical Engineering from Stanford University. He was with the Department of Electrical and Computer Engineering at BYU from 1990-2007, where he served as Department Chair from 2003-06.  During 1996-97, he held a joint appointment as a visiting scholar at Uppsala University and the Royal Institute of Technology in Sweden. From 2006-07, he was on leave working as Vice President of Research for ArrayComm LLC in San Jose, California. Since 2007 he has been with the Electrical Engineering and Computer Science (EECS) Department at the University of California Irvine, where he is a Distinguished Professor and currently serving as Department Chair. Dr. Swindlehurst is a Fellow of the IEEE, during 2014-17 he was also a Hans Fischer Senior Fellow in the Institute for Advanced Studies at the Technical University of Munich, and in 2016, he was elected as a Foreign Member of the Royal Swedish Academy of Engineering Sciences (IVA). He received the 2000 IEEE W. R. G. Baker Prize Paper Award, the 2006 IEEE Communications Society Stephen O. Rice Prize in the Field of Communication Theory, the 2006, 2010 and 2021 IEEE Signal Processing Society's Best Paper Awards, the 2017 IEEE Signal Processing Society Donald G. Fink Overview Paper Award, a Best Paper award at the 2020 and 2024 IEEE International Conferences on Communications, the 2022 Claude Shannon-Harry Nyquist Technical Achievement Award from the IEEE Signal Processing Society, and the 2024 Fred W. Ellersick Prize from the IEEE Communications Society. His research focuses on array signal processing for radar, wireless communications, and biomedical applications.
\end{IEEEbiography}

\end{document}

%% file: abbr.tex
\acrodef{wsts}[WSTS]{WildfireSpreadTS}
\acrodef{ndws}[NDWS]{Next Day Wildfire Spread}
\acrodef{viirs}[VIIRS]{Visible Infrared Imaging Radiometer Suite}
\acrodef{gridmet}[GRIDMET]{Gridded Surface Meteorological}
\acrodef{gfs}[GFS]{Global Forecast System}
\acrodef{modis}[MODIS]{Moderate Resolution Imaging Spectroradiometer}
\acrodef{srtm}[SRTM]{Shuttle Radar Topography Mission}
\acrodef{pdsi}[PDSI]{Palmer Drought Severity Index}
\acrodef{erc}[ERC]{Energy Release Component}
\acrodef{cnn}[CNN]{Convolutional Neural Network}
\acrodef{vit}[ViT]{Vision Transformer}
\acrodef{d3pm}[D3PM]{Discrete Denoising Diffusion Probabilistic Model}
\acrodef{maskgit}[MaskGIT]{Masked Generative Image Transformer}
\acrodef{utae}[U-TAE]{U-net with Temporal Attention Encoder}
\acrodef{convlstm}[ConvLSTM]{convolutional LSTM}
\acrodef{film}[FiLM]{Feature-wise Linear Modulation}
\acrodef{uav}[UAV]{Uncrewed Aerial Vehicle}
\acrodef{sar}[SAR]{Synthetic Aperture Radar}
\acrodef{elbo}[ELBO]{Evidence Lower Bound}
\acrodef{cvae}[CVAE]{Conditional Variational Autoencoder}
\acrodef{kl}[KL]{Kullback-Leibler}
\acrodef{ltae}[L-TAE]{Lightweight Temporal Attention Encoder}
\acrodef{fpr}[FPR]{False Positive Rate}
\acrodef{ap}[AP]{Average Precision}

%% file: tables/table1.tex
\begin{table}[!t]
    \centering
    \captionsetup{justification=justified, singlelinecheck=false}
    \caption{Data statistics for the \textbf{original 23-channel} input dataset \ac{wsts}. Statistics (Min/Max/Mean) are computed globally over all pixels and timesteps across \textbf{607 wildfire events} from \textbf{2018--2021}.}
    \label{table1}
    \tiny
    \setlength{\tabcolsep}{2pt}
    \renewcommand{\arraystretch}{1.1}
    \begin{threeparttable}
        \begin{tabular}{l r l l l r r r}
            \toprule
            \textbf{Category} & \textbf{Ch.} & \textbf{Name} & \textbf{Source} & \textbf{Unit} & \textbf{Min} & \textbf{Max} & \textbf{Mean} \\
            \midrule
            \multirow{16}{*}{\textbf{Measurement}}
              & 1  & Band M11                       & \acs{viirs}      & --          & -100.00     & 16000.00  & 1908.53 \\
              & 2  & Band I2                        & \acs{viirs}      & --          & -100.00     & 15998.00  & 2944.98 \\
              & 3  & Band I1                        & \acs{viirs}      & --          & -100.00     & 15997.00  & 1820.32 \\
              & 4  & NDVI                           & \acs{viirs}      & --          & -9966.00    & 9995.00   & 4323.21 \\
              & 5  & EVI2                           & \acs{viirs}      & --          & -5172.00    & 9998.00   & 2221.04 \\
              & 6  & Total Precipitation            & \acs{gridmet}    & mm          & 0.00        & 145.30    & 0.50    \\
              & 7  & Wind Speed                     & \acs{gridmet}    & m/s         & 0.30        & 16.20     & 3.49    \\
              & 8  & Wind Direction                 & \acs{gridmet}    & $^{\circ}$  & 0.00        & 360.00    & 222.40  \\
              & 9  & Min Temperature                & \acs{gridmet}    & K           & 242.00      & 311.80    & 283.06  \\
              & 10 & Max Temperature                & \acs{gridmet}    & K           & 254.70      & 325.40    & 299.47  \\
              & 11 & Energy Release Component       & \acs{gridmet}    & --          & 0.00        & 122.00    & 68.96   \\
              & 12 & Specific Humidity              & \acs{gridmet}    & kg/kg       & 0.00        & 0.02      & 0.01    \\
              & 13 & Slope                          & \acs{srtm}       & $^{\circ}$  & 0.00        & 67.07     & 7.21    \\
              & 14 & Aspect                         & \acs{srtm}       & $^{\circ}$  & 0.00        & 359.89    & 175.43  \\
              & 15 & Elevation                      & \acs{srtm}       & m           & -84.00      & 4350.00   & 1487.21 \\
              & 16 & Palmer Drought Severity Index  & \acs{gridmet}    & --          & -13.75      & 9.66      & -2.02 \\
            \cmidrule{1-8}
            \textbf{Land Cover}
              & 17 & Land Cover                     & \acs{modis}      & ID          & 1.00        & 17.00     & --      \\
            \cmidrule{1-8}
            \multirow{5}{*}{\textbf{Forecast}}
              & 18 & Total Precipitation            & \acs{gfs}        & mm          & 0.00        & 1144.81   & 9.09    \\
              & 19 & Wind Speed                     & \acs{gfs}        & m/s         & 0.00        & 14.30     & 1.58    \\
              & 20 & Wind Direction                 & \acs{gfs}        & $^{\circ}$  & -89.99      & 89.99     & 6.15    \\
              & 21 & Temperature                    & \acs{gfs}        & $^{\circ}$C & -17.04      & 39.51     & 18.58   \\
              & 22 & Specific Humidity              & \acs{gfs}        & kg/kg       & 0.00        & 0.01      & 0.01    \\
            \cmidrule{1-8}
            \textbf{Fire Map}
              & 23 & Active Fire                    & \acs{viirs}      & HHMM        & 742.00      & 2218.00   & --      \\
            \bottomrule
        \end{tabular}
        \begin{tablenotes}[flushleft]
            \tiny
            \item \textit{Note:} ``--'' denotes unitless or categorical variables. \acs{viirs} active-fire time is in UTC. Measured temperatures are in Kelvin (K); forecasts are in Celsius ($^{\circ}$C). Data sources include \ac{viirs}, \ac{gridmet}, \ac{srtm}, \ac{modis}, and \ac{gfs}.
        \end{tablenotes}
    \end{threeparttable}
\end{table}

%% file: tables/table2.tex
\begin{table}[!t]
\caption{Land cover channel class mapping (ID 1--17).}
\label{table2}
\centering
\footnotesize
\setlength{\tabcolsep}{8pt}
\renewcommand{\arraystretch}{1.15}
\begin{tabular}{r l}
\toprule
\textbf{ID} & \textbf{Land Cover Type} \\
\midrule
1  & Evergreen Needleleaf Forests \\
2  & Evergreen Broadleaf Forests \\
3  & Deciduous Needleleaf Forests \\
4  & Deciduous Broadleaf Forests \\
5  & Mixed Forests \\
6  & Closed Shrublands \\
7  & Open Shrublands \\
8  & Woody Savannas \\
9  & Savannas \\
10 & Grasslands \\
11 & Permanent Wetlands \\
12 & Croplands \\
13 & Urban and Built-up Lands \\
14 & Cropland/Natural Vegetation Mosaics \\
15 & Permanent Snow and Ice \\
16 & Barren \\
17 & Water Bodies \\
\bottomrule
\end{tabular}
\end{table}

%% file: tables/table3.tex
\begin{table*}[t]
\centering
\caption{\textbf{STAGE-1: RECONSTRUCTION} \\ Performance under different fire scenarios (mean $\pm$ std)}
\label{tab:stage1}

\scriptsize
\setlength{\tabcolsep}{2.6pt}
\renewcommand{\arraystretch}{1.1}

\vspace{0.3em}
\textbf{(a) PIXEL-WISE MASKING}
\vspace{0.4em}

\begin{minipage}[t]{0.49\textwidth}
\centering
\textbf{FIRE CONTINUES [DICE]}\\[0.25em]
\begin{tabular}{lcccccc}
\toprule
Mask & Random & Dilation & MaskCVAE & MaskUNet & MaskD3PM & MaskViT \\
\midrule
10\% & 0.055$\pm${\tiny 0.071} & 0.410$\pm${\tiny 0.327} & 0.924$\pm${\tiny 0.126} & 0.920$\pm${\tiny 0.130} & 0.744$\pm${\tiny 0.251} & 0.865$\pm${\tiny 0.184} \\
20\% & 0.055$\pm${\tiny 0.069} & 0.413$\pm${\tiny 0.322} & 0.909$\pm${\tiny 0.116} & 0.906$\pm${\tiny 0.120} & 0.719$\pm${\tiny 0.230} & 0.850$\pm${\tiny 0.169} \\
30\% & 0.055$\pm${\tiny 0.069} & 0.415$\pm${\tiny 0.321} & 0.894$\pm${\tiny 0.115} & 0.890$\pm${\tiny 0.122} & 0.682$\pm${\tiny 0.234} & 0.835$\pm${\tiny 0.163} \\
40\% & 0.055$\pm${\tiny 0.069} & 0.416$\pm${\tiny 0.319} & 0.874$\pm${\tiny 0.125} & 0.870$\pm${\tiny 0.130} & 0.641$\pm${\tiny 0.243} & 0.818$\pm${\tiny 0.165} \\
50\% & 0.055$\pm${\tiny 0.069} & 0.419$\pm${\tiny 0.318} & 0.853$\pm${\tiny 0.135} & 0.849$\pm${\tiny 0.142} & 0.591$\pm${\tiny 0.260} & 0.799$\pm${\tiny 0.170} \\
60\% & 0.055$\pm${\tiny 0.068} & 0.423$\pm${\tiny 0.317} & 0.825$\pm${\tiny 0.155} & 0.821$\pm${\tiny 0.160} & 0.530$\pm${\tiny 0.281} & 0.774$\pm${\tiny 0.183} \\
70\% & 0.055$\pm${\tiny 0.068} & 0.429$\pm${\tiny 0.315} & 0.793$\pm${\tiny 0.179} & 0.789$\pm${\tiny 0.184} & 0.460$\pm${\tiny 0.306} & 0.742$\pm${\tiny 0.202} \\
80\% & 0.055$\pm${\tiny 0.068} & 0.438$\pm${\tiny 0.313} & 0.747$\pm${\tiny 0.216} & 0.743$\pm${\tiny 0.220} & 0.376$\pm${\tiny 0.337} & 0.692$\pm${\tiny 0.233} \\
\bottomrule
\end{tabular}
\end{minipage}\hfill
\begin{minipage}[t]{0.49\textwidth}
\centering
\textbf{FIRE EXTINGUISHED [DICE]}\\[0.25em]
\begin{tabular}{lcccccc}
\toprule
Mask & Random & Dilation & MaskCVAE & MaskUNet & MaskD3PM & MaskViT \\
\midrule
10\% & 0.014$\pm${\tiny 0.035} & 0.614$\pm${\tiny 0.425} & 0.959$\pm${\tiny 0.131} & 0.957$\pm${\tiny 0.134} & 0.877$\pm${\tiny 0.248} & 0.912$\pm${\tiny 0.210} \\
20\% & 0.014$\pm${\tiny 0.035} & 0.615$\pm${\tiny 0.422} & 0.947$\pm${\tiny 0.133} & 0.944$\pm${\tiny 0.139} & 0.852$\pm${\tiny 0.252} & 0.891$\pm${\tiny 0.215} \\
30\% & 0.014$\pm${\tiny 0.034} & 0.616$\pm${\tiny 0.420} & 0.934$\pm${\tiny 0.138} & 0.932$\pm${\tiny 0.143} & 0.825$\pm${\tiny 0.271} & 0.880$\pm${\tiny 0.212} \\
40\% & 0.014$\pm${\tiny 0.034} & 0.618$\pm${\tiny 0.419} & 0.917$\pm${\tiny 0.150} & 0.916$\pm${\tiny 0.152} & 0.791$\pm${\tiny 0.297} & 0.865$\pm${\tiny 0.216} \\
50\% & 0.014$\pm${\tiny 0.034} & 0.618$\pm${\tiny 0.418} & 0.897$\pm${\tiny 0.174} & 0.895$\pm${\tiny 0.178} & 0.754$\pm${\tiny 0.326} & 0.848$\pm${\tiny 0.229} \\
60\% & 0.014$\pm${\tiny 0.034} & 0.620$\pm${\tiny 0.417} & 0.870$\pm${\tiny 0.206} & 0.868$\pm${\tiny 0.208} & 0.712$\pm${\tiny 0.359} & 0.827$\pm${\tiny 0.247} \\
70\% & 0.014$\pm${\tiny 0.034} & 0.621$\pm${\tiny 0.416} & 0.836$\pm${\tiny 0.241} & 0.837$\pm${\tiny 0.242} & 0.667$\pm${\tiny 0.393} & 0.799$\pm${\tiny 0.274} \\
80\% & 0.014$\pm${\tiny 0.034} & 0.624$\pm${\tiny 0.415} & 0.791$\pm${\tiny 0.289} & 0.788$\pm${\tiny 0.294} & 0.617$\pm${\tiny 0.431} & 0.751$\pm${\tiny 0.318} \\
\bottomrule
\end{tabular}
\end{minipage}

\vspace{1.0em}

\textbf{(b) BLOCK-WISE MASKING}
\vspace{0.4em}

\begin{minipage}[t]{0.49\textwidth}
\centering
\textbf{FIRE CONTINUES [DICE]}\\[0.25em]
\begin{tabular}{lcccccc}
\toprule
Mask & Random & Dilation & MaskCVAE & MaskUNet & MaskD3PM & MaskViT \\
\midrule
10\% & 0.052$\pm${\tiny 0.086} & 0.458$\pm${\tiny 0.400} & 0.806$\pm${\tiny 0.276} & 0.800$\pm${\tiny 0.281} & 0.553$\pm${\tiny 0.468} & 0.741$\pm${\tiny 0.323} \\
20\% & 0.054$\pm${\tiny 0.077} & 0.421$\pm${\tiny 0.362} & 0.751$\pm${\tiny 0.280} & 0.745$\pm${\tiny 0.284} & 0.425$\pm${\tiny 0.457} & 0.673$\pm${\tiny 0.319} \\
30\% & 0.055$\pm${\tiny 0.074} & 0.417$\pm${\tiny 0.345} & 0.722$\pm${\tiny 0.277} & 0.716$\pm${\tiny 0.280} & 0.366$\pm${\tiny 0.439} & 0.636$\pm${\tiny 0.311} \\
40\% & 0.055$\pm${\tiny 0.072} & 0.420$\pm${\tiny 0.337} & 0.698$\pm${\tiny 0.279} & 0.692$\pm${\tiny 0.281} & 0.331$\pm${\tiny 0.423} & 0.608$\pm${\tiny 0.307} \\
50\% & 0.055$\pm${\tiny 0.071} & 0.426$\pm${\tiny 0.332} & 0.678$\pm${\tiny 0.280} & 0.672$\pm${\tiny 0.282} & 0.310$\pm${\tiny 0.412} & 0.584$\pm${\tiny 0.306} \\
60\% & 0.055$\pm${\tiny 0.070} & 0.432$\pm${\tiny 0.329} & 0.656$\pm${\tiny 0.283} & 0.649$\pm${\tiny 0.286} & 0.290$\pm${\tiny 0.402} & 0.559$\pm${\tiny 0.305} \\
70\% & 0.055$\pm${\tiny 0.070} & 0.436$\pm${\tiny 0.328} & 0.631$\pm${\tiny 0.291} & 0.623$\pm${\tiny 0.293} & 0.276$\pm${\tiny 0.396} & 0.527$\pm${\tiny 0.309} \\
80\% & 0.055$\pm${\tiny 0.069} & 0.439$\pm${\tiny 0.330} & 0.598$\pm${\tiny 0.301} & 0.590$\pm${\tiny 0.303} & 0.262$\pm${\tiny 0.392} & 0.488$\pm${\tiny 0.315} \\
\bottomrule
\end{tabular}
\end{minipage}\hfill
\begin{minipage}[t]{0.49\textwidth}
\centering
\textbf{FIRE EXTINGUISHED [DICE]}\\[0.25em]
\begin{tabular}{lcccccc}
\toprule
Mask & Random & Dilation & MaskCVAE & MaskUNet & MaskD3PM & MaskViT \\
\midrule
10\% & 0.013$\pm${\tiny 0.043} & 0.698$\pm${\tiny 0.427} & 0.908$\pm${\tiny 0.236} & 0.908$\pm${\tiny 0.237} & 0.837$\pm${\tiny 0.355} & 0.881$\pm${\tiny 0.275} \\
20\% & 0.014$\pm${\tiny 0.039} & 0.635$\pm${\tiny 0.438} & 0.856$\pm${\tiny 0.284} & 0.856$\pm${\tiny 0.284} & 0.756$\pm${\tiny 0.412} & 0.814$\pm${\tiny 0.329} \\
30\% & 0.014$\pm${\tiny 0.037} & 0.617$\pm${\tiny 0.437} & 0.821$\pm${\tiny 0.310} & 0.824$\pm${\tiny 0.307} & 0.710$\pm${\tiny 0.435} & 0.774$\pm${\tiny 0.350} \\
40\% & 0.014$\pm${\tiny 0.036} & 0.613$\pm${\tiny 0.434} & 0.787$\pm${\tiny 0.331} & 0.788$\pm${\tiny 0.331} & 0.667$\pm${\tiny 0.451} & 0.730$\pm${\tiny 0.369} \\
50\% & 0.014$\pm${\tiny 0.035} & 0.613$\pm${\tiny 0.432} & 0.755$\pm${\tiny 0.349} & 0.759$\pm${\tiny 0.346} & 0.641$\pm${\tiny 0.458} & 0.696$\pm${\tiny 0.383} \\
60\% & 0.014$\pm${\tiny 0.035} & 0.615$\pm${\tiny 0.429} & 0.725$\pm${\tiny 0.364} & 0.728$\pm${\tiny 0.362} & 0.619$\pm${\tiny 0.464} & 0.661$\pm${\tiny 0.396} \\
70\% & 0.014$\pm${\tiny 0.035} & 0.616$\pm${\tiny 0.429} & 0.693$\pm${\tiny 0.381} & 0.689$\pm${\tiny 0.382} & 0.596$\pm${\tiny 0.470} & 0.619$\pm${\tiny 0.409} \\
80\% & 0.014$\pm${\tiny 0.034} & 0.615$\pm${\tiny 0.432} & 0.650$\pm${\tiny 0.399} & 0.648$\pm${\tiny 0.400} & 0.580$\pm${\tiny 0.474} & 0.578$\pm${\tiny 0.421} \\
\bottomrule
\end{tabular}
\end{minipage}

\vspace{0.8em}

\end{table*}

%% file: tables/table5.tex
\begin{table}[t]
\centering
\caption{\textbf{STAGE-2: FORECASTING} \\ UTAE Performance under different fire scenarios (mean $\pm$ std).}
\label{tab:stage2}
\scriptsize
\setlength{\tabcolsep}{2pt}
\renewcommand{\arraystretch}{1.1}
\begin{tabular}{c cc cc}
\toprule
 & \multicolumn{2}{c}{\textbf{PIXEL-WISE MASKING}} & \multicolumn{2}{c}{\textbf{BLOCK-WISE MASKING}} \\
\cmidrule(lr){2-3} \cmidrule(lr){4-5}
Mask & FIRE CONT. [AP] & FIRE EXT. [FPR] & FIRE CONT. [AP] & FIRE EXT. [FPR] \\
\midrule
0\%  & 0.527$\pm${\tiny 0.274} & 0.002$\pm${\tiny 0.010} & 0.527$\pm${\tiny 0.273} & 0.002$\pm${\tiny 0.010} \\
10\% & 0.521$\pm${\tiny 0.274} & 0.002$\pm${\tiny 0.010} & 0.508$\pm${\tiny 0.269} & 0.002$\pm${\tiny 0.010} \\
20\% & 0.513$\pm${\tiny 0.274} & 0.002$\pm${\tiny 0.009} & 0.487$\pm${\tiny 0.267} & 0.001$\pm${\tiny 0.009} \\
30\% & 0.501$\pm${\tiny 0.275} & 0.001$\pm${\tiny 0.008} & 0.465$\pm${\tiny 0.261} & 0.001$\pm${\tiny 0.008} \\
40\% & 0.491$\pm${\tiny 0.274} & 0.001$\pm${\tiny 0.007} & 0.445$\pm${\tiny 0.260} & 0.001$\pm${\tiny 0.007} \\
50\% & 0.475$\pm${\tiny 0.273} & 0.001$\pm${\tiny 0.006} & 0.418$\pm${\tiny 0.256} & 0.001$\pm${\tiny 0.006} \\
60\% & 0.454$\pm${\tiny 0.272} & 0.001$\pm${\tiny 0.005} & 0.396$\pm${\tiny 0.251} & 0.001$\pm${\tiny 0.006} \\
70\% & 0.426$\pm${\tiny 0.269} & 0.000$\pm${\tiny 0.004} & 0.369$\pm${\tiny 0.247} & 0.001$\pm${\tiny 0.005} \\
80\% & 0.385$\pm${\tiny 0.264} & 0.000$\pm${\tiny 0.003} & 0.334$\pm${\tiny 0.241} & 0.000$\pm${\tiny 0.004} \\
\bottomrule
\end{tabular}
\end{table}

%% file: ref.bib
@inproceedings{Gerard2023WildfireSpreadTS,
  author    = {S. Gerard and Y. Zhao and J. Sullivan},
  title     = {{WildfireSpreadTS}: A Dataset of Multi-Modal Time Series for Wildfire Spread Prediction},
  booktitle = {Proc. Adv. Neural Inf. Process. Syst. (NeurIPS)},
  volume    = {36},
  pages     = {74515--74529},
  year      = {2023}
}

@article{Schroeder2014VIIRS,
  author  = {W. Schroeder and P. Oliva and L. Giglio and I. A. Csiszar},
  title   = {The New {VIIRS} 375 m Active Fire Detection Data Product: Algorithm Description and Initial Assessment},
  journal = {Remote Sens. Environ.},
  volume  = {143},
  pages   = {85--96},
  year    = {2014},
  doi     = {10.1016/j.rse.2013.12.008}
}

@article{Abatzoglou2013GriddedMet,
  author  = {J. T. Abatzoglou},
  title   = {Development of gridded surface meteorological data for ecological applications and modelling},
  journal = {Int. J. Climatol.},
  volume  = {33},
  number  = {1},
  pages   = {121--131},
  year    = {2013},
  doi     = {10.1002/joc.3413}
}

@article{Oliva2015VIIRSBA,
  author  = {P. Oliva and W. Schroeder},
  title   = {Assessment of {VIIRS} 375m Active Fire Detection Product for Direct Burned Area Mapping},
  journal = {Remote Sens. Environ.},
  volume  = {160},
  pages   = {144--155},
  year    = {2015},
  doi     = {10.1016/j.rse.2015.01.010}
}

@misc{Vermote2016VNP09GA,
  author       = {E. Vermote and others},
  title        = {{VIIRS}/{NPP} Surface Reflectance Daily {L2G} Global 1km and 500m {SIN} Grid {V001} [Data set]},
  howpublished = {NASA Land Processes Distributed Active Archive Center (LP DAAC)},
  year         = {2016},
  doi          = {10.5067/VIIRS/VNP09GA.001}
}

@misc{Didan2018VNP13A1,
  author       = {K. Didan and A. Barreto},
  title        = {{VIIRS}/{NPP} Vegetation Indices 16-Day {L3} Global 500 m {SIN} Grid {V001}},
  howpublished = {Dataset, NASA Land Processes Distributed Active Archive Center (LP DAAC)},
  year         = {2018},
  doi          = {10.5067/VIIRS/VNP13A1.001},
  note         = {Accessed: Dec. 27, 2025}
}

@article{Clough2005AER,
  author  = {S. A. Clough and M. W. Shephard and E. J. Mlawer and J. S. Delamere and M. J. Iacono and K. Cady-Pereira and S. Boukabara and P. D. Brown},
  title   = {Atmospheric Radiative Transfer Modeling: A Summary of the {AER} Codes},
  journal = {J. Quant. Spectrosc. Radiat. Transfer},
  volume  = {91},
  number  = {2},
  pages   = {233--244},
  year    = {2005},
  doi     = {10.1016/j.jqsrt.2004.05.058}
}

@article{SullaMenashe2019MCD12Q1C6,
  author  = {D. Sulla-Menashe and J. M. Gray and S. P. Abercrombie and M. A. Friedl},
  title   = {Hierarchical Mapping of Annual Global Land Cover 2001 to Present: The {MODIS} {C}ollection 6 {Land Cover Product}},
  journal = {Remote Sens. Environ.},
  volume  = {222},
  pages   = {183--194},
  year    = {2019},
  doi     = {10.1016/j.rse.2018.12.013}
}

@misc{FriedlSullaMenashe_MCD12Q1_061_2022,
  author       = {Friedl, Mark and Sulla-Menashe, Deborah},
  title        = {{MODIS/Terra+Aqua Land Cover Type Yearly L3 Global 500m SIN Grid V061}},
  howpublished = {Dataset, NASA Land Processes Distributed Active Archive Center (LP DAAC)},
  year         = {2022},
  doi          = {10.5067/MODIS/MCD12Q1.061},
  note         = {Accessed: 2026-01-01}
}

@misc{NASA2013SRTMGL1,
  author       = {{NASA JPL}},
  title        = {{NASA} Shuttle Radar Topography Mission Global 1 Arc Second [Data set]},
  howpublished = {NASA Land Processes Distributed Active Archive Center (LP DAAC)},
  year         = {2013},
  doi          = {10.5067/MEASURES/SRTM/SRTMGL1.003},
  note         = {Accessed: Dec. 28, 2025}
}

@inproceedings{Ronneberger2015UNet,
  author    = {Ronneberger, Olaf and Fischer, Philipp and Brox, Thomas},
  title     = {U-Net: Convolutional Networks for Biomedical Image Segmentation},
  booktitle = {Med. Image Comput. Comput.-Assist. Intervent.---MICCAI 2015},
  editor    = {Navab, Nassir and Hornegger, Joachim and Wells, William M. and Frangi, Alejandro F.},
  series    = {Lect. Notes Comput. Sci.},
  volume    = {9351},
  pages     = {234--241},
  address   = {Cham, Switzerland},
  publisher = {Springer},
  year      = {2015},
  doi       = {10.1007/978-3-319-24574-4_28}
}

@techreport{sullivan2022spreading,
  author      = {Andrew Sullivan and Elaine Baker and Tiina Kurvits and Alexandra Popescu and Alison K. Paulson and Amy Cardinal Christianson and Ayesha Tulloch and Bibiana Bilbao and Camilla Mathison and Catherine Robinson and others},
  title       = {Spreading like Wildfire: The Rising Threat of Extraordinary Landscape Fires},
  institution = {{United Nations Environment Programme}},
  year        = {2022},
  type        = {Report}
}

@article{law2025anthropogenic,
  author  = {B. E. Law and J. T. Abatzoglou and C. R. Schwalm and D. Byrne and N. Fann and N. J. Nassikas},
  title   = {Anthropogenic Climate Change Contributes to Wildfire Particulate Matter and Related Mortality in the {United States}},
  journal = {Commun. Earth Environ.},
  volume  = {6},
  number  = {1},
  artno   = {336},
  year    = {2025}
}

@article{dickman2021ecological,
  author  = {Christopher R. Dickman},
  title   = {Ecological Consequences of Australia's ``Black Summer'' Bushfires: Managing for Recovery},
  journal = {Integr. Environ. Assess. Manag.},
  volume  = {17},
  number  = {6},
  pages   = {1162--1167},
  year    = {2021}
}

@article{reiner2024_black_summer_tourism,
  author  = {Vivienne Reiner and Navoda Liyana Pathirana and Ya-Yen Sun and Manfred Lenzen and Arunima Malik},
  title   = {Wish You Were Here? {The} Economic Impact of the Tourism Shutdown from {A}ustralia's 2019--20 {``Black Summer''} Bushfires},
  journal = {Econ. Disaster Clim. Chang.},
  volume  = {8},
  number  = {1},
  pages   = {107--127},
  month   = mar,
  year    = {2024},
  doi     = {10.1007/s41885-024-00142-8}
}

@article{paglino2025excess,
  author  = {E. Paglino and R. V. Raquib and A. C. Stokes},
  title   = {Excess Deaths Attributable to the {Los Angeles} Wildfires from {Jan}.~5 to {Feb}.~1, 2025},
  journal = {JAMA},
  volume  = {334},
  number  = {11},
  pages   = {1018--1019},
  year    = {2025}
}

@article{chan2024survey,
  author  = {C.-C. Chan and S. A. Alvi and X. Zhou and S. Durrani and N. Wilson and M. Yebra},
  title   = {A Survey on {IoT} Ground Sensing Systems for Early Wildfire Detection: Technologies, Challenges, and Opportunities},
  journal = {IEEE Access},
  volume  = {12},
  pages   = {172785--172819},
  year    = {2024}
}

@article{justice2002modis,
  author  = {C. O. Justice and L. Giglio and S. Korontzi and J. Owens and J. T. Morisette and D. Roy and J. Descloitres and S. Alleaume and F. Petitcolin and Y. Kaufman},
  title   = {The {MODIS} Fire Products},
  journal = {Remote Sens. Environ.},
  volume  = {83},
  number  = {1--2},
  pages   = {244--262},
  year    = {2002}
}

@article{bouguettaya2022review,
  author  = {A. Bouguettaya and H. Zarzour and A. M. Taberkit and A. Kechida},
  title   = {A Review on Early Wildfire Detection from Unmanned Aerial Vehicles Using Deep Learning-Based Computer Vision Algorithms},
  journal = {Signal Process.},
  volume  = {190},
  artno   = {108309},
  year    = {2022}
}

@article{ghali2023deep,
  author  = {R. Ghali and M. A. Akhloufi},
  title   = {Deep Learning Approaches for Wildland Fires Using Satellite Remote Sensing Data: Detection, Mapping, and Prediction},
  journal = {Fire},
  volume  = {6},
  number  = {5},
  artno   = {192},
  year    = {2023}
}

@article{andrianarivony2024machine,
  author  = {H. S. Andrianarivony and M. A. Akhloufi},
  title   = {Machine Learning and Deep Learning for Wildfire Spread Prediction: A Review},
  journal = {Fire},
  volume  = {7},
  number  = {12},
  artno   = {482},
  year    = {2024}
}

@inproceedings{radke2019firecast,
  author    = {D. Radke and A. Hessler and D. Ellsworth},
  title     = {{FireCast}: Leveraging Deep Learning to Predict Wildfire Spread},
  booktitle = {Proc. Int. Joint Conf. Artif. Intell. (IJCAI)},
  pages     = {4575--4581},
  year      = {2019}
}

@article{huot2022next,
  author  = {F. Huot and R. L. Hu and N. Goyal and T. Sankar and M. Ihme and Y.-F. Chen},
  title   = {Next Day Wildfire Spread: A Machine Learning Dataset to Predict Wildfire Spreading from Remote-Sensing Data},
  journal = {IEEE Trans. Geosci. Remote Sens.},
  volume  = {60},
  pages   = {1--13},
  year    = {2022}
}

@article{schmit2017closer,
  author  = {T. J. Schmit and P. Griffith and M. M. Gunshor and J. M. Daniels and S. J. Goodman and W. J. Lebair},
  title   = {A Closer Look at the {ABI} on the {GOES-R} Series},
  journal = {Bull. Amer. Meteor. Soc.},
  volume  = {98},
  number  = {4},
  pages   = {681--698},
  year    = {2017}
}

@article{savtchenko2004terra,
  author  = {A. Savtchenko and D. Ouzounov and S. Ahmad and J. Acker and G. Leptoukh and J. Koziana and D. Nickless},
  title   = {Terra and Aqua {MODIS} products available from {NASA} {GES} {DAAC}},
  journal = {Adv. Space Res.},
  year    = {2004},
  vol.    = {34},
  no.     = {4},
  pp.     = {710--714},
  doi     = {10.1016/j.asr.2004.03.012}
}

@article{chuvieco2020satellite,
  author  = {E. Chuvieco and I. Aguado and J. Salas and M. Garc{\'i}a and M. Yebra and P. Oliva},
  title   = {Satellite Remote Sensing Contributions to Wildland Fire Science and Management},
  journal = {Curr. Forestry Rep.},
  year    = {2020},
  vol.    = {6},
  no.     = {2},
  pp.     = {81--96},
  doi     = {10.1007/s40725-020-00116-5}
}

@article{khennou2023improving,
  author  = {F. Khennou and M. A. Akhloufi},
  title   = {Improving Wildland Fire Spread Prediction Using Deep {U}-{N}ets},
  journal = {Sci. Remote Sens.},
  year    = {2023},
  vol.    = {8},
  pages   = {100101},
  doi     = {10.1016/j.srs.2023.100101}
}

@inproceedings{li2024wildfirevit,
  author    = {Li, Bronte Sihan and Rad, Ryan},
  title     = {Wildfire Spread Prediction in {North America} Using Satellite Imagery and {V}ision {T}ransformer},
  booktitle = {Proc. 2024 {IEEE} Conf. on Artificial Intelligence ({CAI})},
  year      = {2024},
  pages     = {1536--1541},
  doi       = {10.1109/CAI59869.2024.00278},
  publisher = {IEEE}
}

@article{zhao2025ts,
  author  = {Y. Zhao and S. Gerard and Y. Ban},
  title   = {{TS-SatFire}: A Multi-Task Satellite Image Time-Series Dataset for Wildfire Detection and Prediction},
  journal = {Sci. Data},
  year    = {2025},
  vol.    = {12},
  no.     = {1},
  pages   = {1817},
  doi     = {10.1038/s41597-025-06271-3}
}

@inproceedings{shi2015convlstm,
  author    = {Shi, Xingjian and Chen, Zhourong and Wang, Hao and Yeung, Dit{-}Yan and Wong, Wai{-}Kin and Woo, Wang{-}chun},
  title     = {Convolutional {LSTM} Network: A Machine Learning Approach for Precipitation Nowcasting},
  booktitle = {Adv. Neural Inf. Process. Syst.},
  editor    = {Cortes, Corinna and Lawrence, Neil D. and Lee, Daniel D. and Sugiyama, Masashi and Garnett, Roman},
  volume    = {28},
  pages     = {802--810},
  year      = {2015}
}

@inproceedings{sykas2023eo4wildfires,
  author    = {D. Sykas and D. Zografakis and K. Demestichas and C. Costopoulou and P. Kosmidis},
  title     = {{EO4WildFires}: An Earth Observation Multi-Sensor, Time-Series Machine-Learning-Ready Benchmark Dataset for Wildfire Impact Prediction},
  booktitle = {Proc. SPIE, 9th Int. Conf. Remote Sens. Geoinf. Environ. (RSCy 2023)},
  year      = {2023},
  vol.      = {12786},
  pages     = {1278603},
  doi       = {10.1117/12.2680777},
  publisher = {SPIE}
}

@inproceedings{dosovitskiy2020vit,
  author    = {A. Dosovitskiy and L. Beyer and A. Kolesnikov and D. Weissenborn and X. Zhai and T. Unterthiner and M. Dehghani and M. Minderer and G. Heigold and S. Gelly and J. Uszkoreit and N. Houlsby},
  title     = {An Image is Worth 16x16 Words: Transformers for Image Recognition at Scale},
  booktitle = {Proc. Int. Conf. Learn. Represent. (ICLR)},
  year      = {2021}
}

@inproceedings{kondylatos2023mesogeos,
  author    = {Kondylatos, Spyridon and Prapas, Ioannis and Camps-Valls, Gustau and Papoutsis, Ioannis},
  title     = {{Mesogeos}: A Multi-Purpose Dataset for Data-Driven Wildfire Modeling in the {M}editerranean},
  booktitle = {Adv. Neural Inf. Process. Syst.},
  volume    = {36},
  pages     = {50661--50676},
  year      = {2023}
}

@inproceedings{vaswani2017attention,
  author    = {Vaswani, Ashish and Shazeer, Noam and Parmar, Niki and Uszkoreit, Jakob and Jones, Llion and Gomez, Aidan N. and Kaiser, {\L}ukasz and Polosukhin, Illia},
  title     = {Attention Is All You Need},
  booktitle = {Adv. Neural Inf. Process. Syst.},
  volume    = {30},
  pages     = {5998--6008},
  year      = {2017}
}

@article{schroeder2008quantifying,
  author  = {Schroeder, W. and Csiszar, I. and Morisette, J.},
  title   = {Quantifying the Impact of Cloud Obscuration on Remote Sensing of Active Fires in the {Brazilian} {A}mazon},
  journal = {Remote Sensing of Environment},
  year    = {2008},
  volume  = {112},
  number  = {2},
  pages   = {456--470},
  doi     = {10.1016/j.rse.2007.05.004}
}

@article{giglio2016collection,
  author  = {Giglio, L. and Schroeder, W. and Justice, C. O.},
  title   = {The Collection 6 {MODIS} Active Fire Detection Algorithm and Fire Products},
  journal = {Remote Sensing of Environment},
  year    = {2016},
  volume  = {178},
  pages   = {31--41},
  doi     = {10.1016/j.rse.2016.02.054}
}

@inproceedings{sohn2015cvae,
  author    = {Sohn, Kihyuk and Lee, Honglak and Yan, Xinchen},
  title     = {Learning Structured Output Representation using Deep Conditional Generative Models},
  booktitle = {Adv. Neural Inf. Process. Syst.},
  volume    = {28},
  pages     = {3483--3491},
  year      = {2015}
}

@inproceedings{kingma2013vae,
  author    = {D. P. Kingma and M. Welling},
  title     = {Auto-Encoding Variational Bayes},
  booktitle = {Proc. Int. Conf. Learn. Represent. (ICLR)},
  year      = {2014}
}

@inproceedings{he2016resnet,
  author    = {He, Kaiming and Zhang, Xiangyu and Ren, Shaoqing and Sun, Jian},
  title     = {Deep Residual Learning for Image Recognition},
  booktitle = {Proc. {IEEE} Conf. on Computer Vision and Pattern Recognition ({CVPR})},
  year      = {2016},
  month     = jun,
  pages     = {770--778},
  doi       = {10.1109/CVPR.2016.90}
}

@inproceedings{austin2021structured,
  author    = {Austin, Jacob and Johnson, Daniel D. and Ho, Jonathan and Tarlow, Daniel and Van Den Berg, Rianne},
  title     = {Structured Denoising Diffusion Models in Discrete State-Spaces},
  booktitle = {Adv. Neural Inf. Process. Syst.},
  volume    = {34},
  pages     = {17981--17993},
  year      = {2021}
}

@inproceedings{ho2020denoising,
  author    = {J. Ho and A. Jain and P. Abbeel},
  title     = {Denoising diffusion probabilistic models},
  booktitle = {Adv. Neural Inf. Process. Syst.},
  volume    = {33},
  pages     = {6840--6851},
  year      = {2020}
}

@inproceedings{garnot2020lightweight,
  author    = {V. S. F. Garnot and L. Landrieu},
  title     = {Lightweight temporal self-attention for classifying satellite images time series},
  booktitle = {Int. Workshop Adv. Anal. Learn. Temporal Data},
  pages     = {171--181},
  publisher = {Springer},
  year      = {2020}
}

@inproceedings{garnot2021panoptic,
  title={Panoptic segmentation of satellite image time series with convolutional temporal attention networks},
  author={Garnot, Vivien Sainte Fare and Landrieu, Loic},
  booktitle={Proc. IEEE/CVF Int'l Conf. on Computer Vision},
  pages={4872--4881},
  year={2021}
}

@misc{githubrepo,
    author = {Chen Yang and Mehdi Zafari and Ziheng Duan},
    title = {{Robust-Wildfire-Forecasting}},
    year = {2026},
    howpublished = {GitHub repository},
    url = {https://github.com/LS-Wireless/Robust-Wildfire-Forecasting},
    note = {Accessed: March 8, 2026},
    version = {1.0.0}
}
